\documentclass[aps,preprint,showpacs,floats,epsf,epsfig,nofootinbib,12pt]{revtex4}
\usepackage{graphicx}% Include figure files
\usepackage{epsfig}
\usepackage[dvips]{color}
\usepackage{subfigure}

\def\beq{\begin{eqnarray}}
\def\eeq{\end{eqnarray}}

\begin{document}

\title{Study on possible molecular states composed of $\Lambda_c\bar D$ ($\Lambda_b B$) and $\Sigma_c\bar D$ ($\Sigma_b B$)   within the Bethe-Salpeter framework }

\vspace{1cm}

\author{ Hong-Wei Ke$^{1}$   \footnote{Corresponding author, khw020056@tju.edu.cn}, Mei Li$^{1}$, Xiao-Hai Liu$^{1}$ \footnote{xiaohai.liu@tju.edu.cn}and
        Xue-Qian Li$^2$\footnote{lixq@nankai.edu.cn},
   }

\affiliation{  $^{1}$ School of Science, Tianjin University,
Tianjin 300072, China
\\
  $^{2}$ School of Physics, Nankai University, Tianjin 300071, China }

\vspace{12cm}

\begin{abstract}

$P_c(4312)$ observed by the LHCb collaboration is confirmed as a
pentaquark and its structure, production, and decay behaviors
attract great attention from theorists and experimentalists. Since
its mass is very close to sum of $\Sigma_c$ and $\bar D$ masses,
it is naturally tempted to be considered as a molecular state
composed of $\Sigma_c$ and $\bar D$. Moreover, $P_c(4312)$ is
observed in the channel with $J/\psi p$ final state, requiring
that isospin conservation $P_c(4312)$ is an isospin-1/2
eigenstate. In literature, several groups used various models to
estimate its spectrum. We systematically study the pentaquarks
within the framework of the Bethe-Salpeter equation; thus
$P_c(4312)$ is an excellent target because of the available data.
We calculate the spectrum of $P_c(4312)$ in terms of the
Bethe-Salpter equations and further study its decay modes. Some
predictions on other possible pentaquark states that can be tested
in future experiments are made.


\end{abstract}

\maketitle

\section{Introduction}

Due to the innovation of experimental techniques and facilities as
well as the advances in theory of recent years several exotic
states have been experimentally observed and theoretically
studied. Indeed, more constituents would cause more ambiguities,
unlike the simplest $q\bar q$ for mesons and $qqq$ for baryons.
The inner structures of the exotic states are still not clear yet,
those discoveries stir up large numbers of
discussions\cite{Chen:2016spr}. Indeed the theoretical exploration
is crucial for getting a better understanding of the quark model
and obtaining valuable information about non-perturbative physics.
Definitely, to complete the theoretical job achieving more
accurate data would compose the key.

Some hidden charm or bottom states were measured in two-meson
final
states\cite{Choi:2003ue,Abe:2007jn,Choi:2005,Choi:2007wga,Aubert:2005rm,
Ablikim:2013emm,Ablikim:2013wzq,Ablikim:2013mio,Liu:2013dau,Collaboration:2011gja}.
They are regarded as tetraquark states or meson-meson molecular
states. In 2003 a baryon was measured by LEPS\cite{Nakano:2003qx}
which was conjectured as a pentaquark, however later the
allegation was negated by further more accurate experiments.
Breaking the frustration on existence of pentaquark which was
predicted by Gell-Mann in his first paper on quark model, the LHCb
collaboration reported two pentaquark states observed in
$\Lambda_b$ decays where peaks appear at the $J/\psi p$ final
states\cite{Aaij:2015tga}.

Recently another narrow pentaquark state
$P_c(4312)$\cite{Aaij:2019vzc} has also been observed in the
$J/\psi p$ mass spectrum.  Its mass and width are
$4311.9\pm0.7^{+6.8}_{-0.6}$ MeV and $9.8\pm 2.7^{+3.7}_{-4.5}$
MeV respectively. Since its mass is very close to the sum of
$\Sigma_c$ and $\bar D$ masses, it is natural to regard it as a
molecular state of $\Sigma_c \bar
D$\cite{Chen:2019bip,Liu:2019tjn,Xiao:2019aya,Liu:2019zvb,He:2019ify,Xiao:2019mst,Zhang:2019xtu,Wang:2019hyc,Xu:2019zme,
Wang:2019got,Chen:2019asm,Lin:2019qiv}. Furthermore its width is
rather wide in accordance with the property of molecular states,
so the phenomenon further supports the proposal of its molecular
structure. Some other theorists conjecture $P_c(4312)$ as a
compact pentaquark.\cite{Cheng:2019obk,Wang:2019got} instead. In
Ref. \cite{Fernandez-Ramirez:2019koa} the authors think the
interaction between $\Sigma_c $ and $\bar D$ is too weak to bind
them into a bound state. It is worth deeper explorations about
whether the molecule picture is reasonable. In this work we will
calculate the mass spectrum of $P_c(4312)$ based on the assumption
that it is a stable bound state of $\Sigma_c$ and $\bar D$.
Additionally we also study other possible bound states of
$\Lambda_c\bar D$, $\Lambda_b B$ and $\Sigma_b B$ and see if they
can be formed.

We will employ the Bethe-Salpeter (B-S) equation to study the
possible bound state which consists of a baryon and a meson. The
B-S equation is a relativistic equation to deal with the bound
state and established on the basis of quantum field
theory\cite{Salpeter:1952ib}. Initially, people used the B-S
equation to study the bound state of two
fermions\cite{Chang:2004im,Chang:2005sd} and the system of
one-fermion-one-boson\cite{Guo:1998ef}. In
Ref.\cite{Guo:2007mm,Feng:2011zzb} the authors employed the
Bethe-Salpeter equation to study the $K\bar K$ and $B\bar K$
molecular states and their decays. With the same approach we
studied the molecular state of $B\pi$\cite{Ke:2018jql},
$D^{(*)}D^{(*)}$ and $B^{(*)}B^{(*)}$\cite{Ke:2012gm}. Recently
the approach is extended to explore double charmed
baryons\cite{Weng:2010rb,Li:2019ekr} and pentaquarks which are
assumed to be two-body bound systems. In Ref.\cite{Wang:2019krq}
the authors studied possible bound states of $\Lambda$ ($\Sigma$)
and $\bar K$. In this work we will employ a similar approach to
study the possible bound states of $\Sigma_c\bar D$,
$\Lambda_c\bar D$, $\Lambda_b B$ and $\Sigma_b B$.

At present, pentaquark states $P_c(4312)$, $P_c(4380)$,
$P_c(4440)$ and $P_c(4457)$ have been measured in decays of
$\Lambda_b$ where the pentaquark states peak up at the invariant
mass spectrum of $J/\psi p$, so their isospin is $\frac{1}{2}$
because of isospin conservation. Thus we require that the two
hadron constituents reside in an isospin eigenstate. Instead, for
the $\Lambda_c\bar D$ (as well $\Lambda_b B$) system  its isospin
must be $\frac{1}{2}$ but the $\Sigma_c\bar D$ ( or $\Sigma_b B$)
system may reside in either isospin $\frac{1}{2}$ or
$\frac{3}{2}$. Certainly, for  a bound system with  spin-parity
$\frac{1}{2}^-$ the two constituents are in the $S$wave.

For carrying on our calculation the interactions between two
constituents are needed. According to the quantum field theory two
particles interact via exchanging certain mediate particles. Since
two constituents in a pentaquark are color-singlet hadrons the
exchanged particles are some light hadrons such as $\rho$ or (and)
$\omega$ etc.. The effective interactions are deduced from the
chiral Lagrangian\cite{Ronchen:2012eg,Shen:2016tzq,He:2017aps}
which we list in Appendix A. With the effective interactions we
obtain the kernel and establish the corresponding B-S equation.

With a reasonable parameter set, the B-S equation is solved. For a
spin-isospin eigenstate, if the equation does not possess a
solution, then we would conclude that the corresponding bound
state should not exist in nature; on the contrary, a solution of
the B-S equation implies the bound state being formed. At the same
time the B-S wave function is obtained and  we are able to use the
corresponding formula for calculating the rates of strong decay
$P_c(4312)\to {\rm proton}+\mathcal{V}$ (vector) which can be
compared with the data.

This paper is organized as follows: after this introduction we
will derive the B-S equations related to  possible bound states
composed of a baryon and a meson and the formula for its strong decays.
Then in section III we will solve the B-S equation numerically and
present our  results by figures and tables. Section IV is devoted to a brief
summary.

\section{The bound states of $\Lambda_c\bar D$  and
$\Sigma_c\bar D$ }
\begin{figure*}
        \centering
        \subfigure[~]{
          \includegraphics[width=7cm]{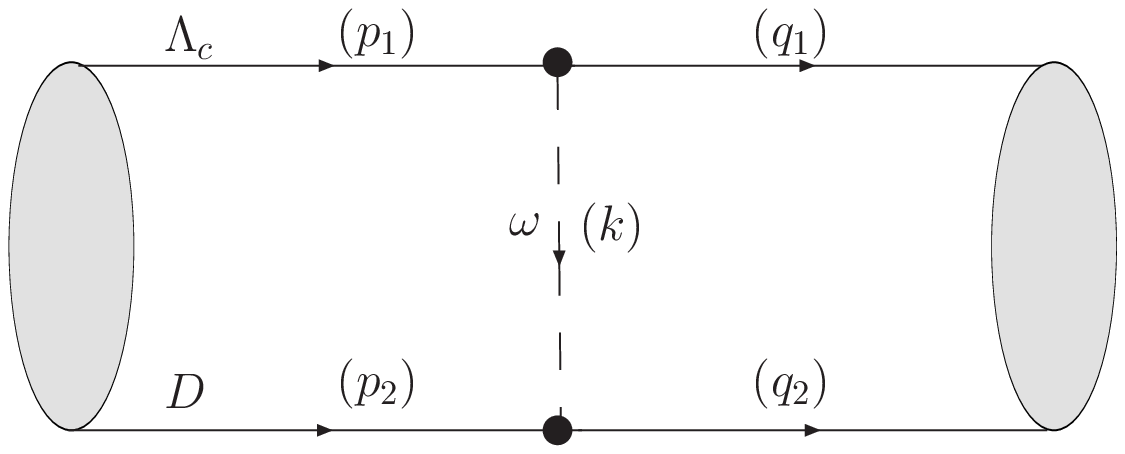}}
        \subfigure[~]{
          \includegraphics[width=7cm]{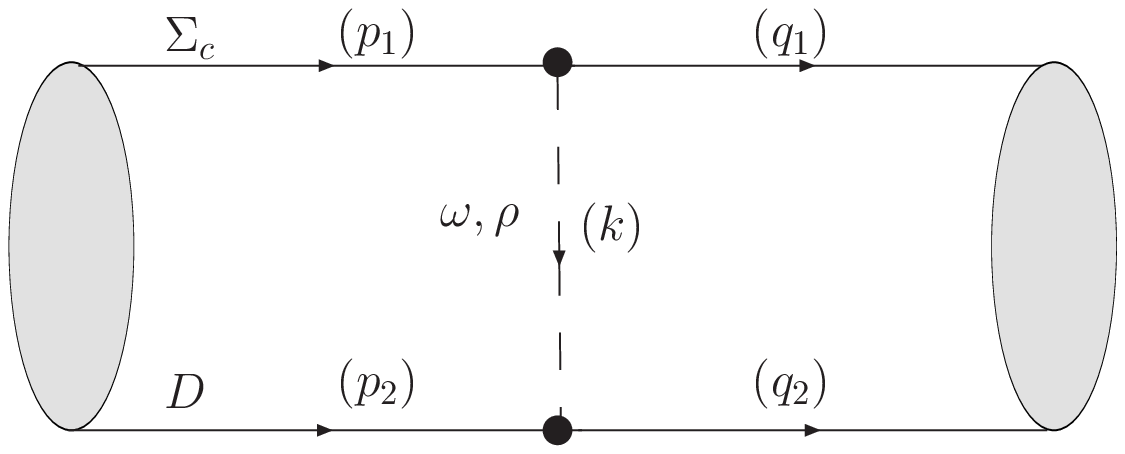}}
 \caption{the bound states of  $\Lambda_c \bar D$ (a) and $\Sigma_c \bar D$ (b)formed by exchanging light vector meson(s) .}
        \label{penta1}
    \end{figure*}

Since the newly found pentaquarks  $P_c(4312)$, $P_c(4380),
P_c(4440)$ and $P_c(4457)$ are all  hadrons containing hidden
charms(or hidden bottoms) and their masses are close to the sums
of the masses of several real hadrons, we will focus on the
molecular structures composed of one charmed (bottomed) baryon and
an anti-charmed(anti-bottomed) meson. Concretely, in this paper we
study  $\Lambda_c\bar D$, $\Sigma_c\bar D$, $\Lambda_b B$ and
$\Sigma_b B$ systems whose spin-parity is $\frac{1}{2}^-$ i.e. the
spatial wave function is in $S$wave. In this section as an example
we only formulate the corresponding quantities for $\Lambda_c\bar
D$ and $\Sigma_c\bar D$ systems. These formulas can be equally
applied to $\Lambda_b B$ and $\Sigma_b B$ systems.

\subsection{The isospin states of $\Lambda_c\bar D$ and $\Sigma_c\bar D$}

The isospin structure of the possible bound state of
$\Lambda_c\bar D$ is
\begin{eqnarray}  \label{isospin1}
|\frac{1}{2},\frac{1}{2}\rangle=|\Lambda_c\bar D^0\rangle.
\end{eqnarray}
We will use $P'_{c(\frac{1}{2},\frac{1}{2})}$ to denote this resonance.

Instead, the possible bound states of
$\Sigma_c\bar D$ should be in three isospin assignments i.e.
$|I,I_3\rangle$ are $|\frac{1}{2},\pm\frac{1}{2}\rangle$
 $|\frac{3}{2},\pm\frac{1}{2}\rangle$ and $|\frac{3}{2},\pm\frac{3}{2}\rangle$. Let us work out the explicit
isospin states
\begin{eqnarray}  \label{isospin2}
|\frac{1}{2},\frac{1}{2}\rangle=\sqrt{\frac{2}{3}}|\Sigma_c^{++}
D^-\rangle-\sqrt{\frac{1}{3}}|\Sigma_c^+\bar D^0\rangle,
\end{eqnarray}
\begin{eqnarray}  \label{isospin3}
|\frac{3}{2},\frac{1}{2}\rangle=\sqrt{\frac{1}{3}}|\Sigma_c^{++}
D^-\rangle+\sqrt{\frac{2}{3}}|\Sigma_c^+\bar D^0\rangle,
\end{eqnarray}
and
\begin{eqnarray}  \label{isospin4}
|\frac{3}{2},\frac{3}{2}\rangle=|\Sigma_c^{++}\bar D^0\rangle.
\end{eqnarray}
The states $|\frac{1}{2},-\frac{1}{2}\rangle$,
$|\frac{3}{2},-\frac{1}{2}\rangle$  and
$|\frac{3}{2},-\frac{3}{2}\rangle$ are just the charge conjugate
states of  $|\frac{1}{2},\frac{1}{2}\rangle$,
$|\frac{3}{2},\frac{1}{2}\rangle$  and
$|\frac{3}{2},\frac{1}{2}\rangle$, therefore their hadronic
properties are the same. We use $P_{c(\frac{1}{2},\frac{1}{2})}$,
$P_{c(\frac{3}{2},\frac{1}{2})}$ and
$P_{c(\frac{3}{2},\frac{3}{2})}$ to denote the three isospin
states of $\Sigma_c\bar D$: $|\frac{1}{2},\frac{1}{2}\rangle$,
$|\frac{3}{2},\frac{1}{2}\rangle$  and
$|\frac{3}{2},\frac{3}{2}\rangle$  respectively for latter
discussions.

In order to discuss the Isospin factors in the B-S equation  we
define the fields of baryons and mesons  in the
expressions\cite{Wang:2019krq}:
\begin{eqnarray}  \label{field}
&&\mathcal{B}_1(x)=\int{d^4{q}\over(2\pi)^4\sqrt{2m_\mathcal{B^{++}}}}
(a_{\mathcal{B}^{--}}e^{-iqx}+a^\dagger_{\mathcal{B}^{++}}e^{iqx}),\nonumber\\&&
\mathcal{B}_2(x)=\int{d^4{q}\over(2\pi)^4\sqrt{2m_\mathcal{B^{+}}}}
(a_{\mathcal{B}^{-}}e^{-iqx}+a^\dagger_{\mathcal{B}^{+}}e^{iqx}),\nonumber\\&&
\mathcal{M}_1(x)=\int{d^4{q}\over(2\pi)^4\sqrt{2m_\mathcal{M^{+}}}}
(a_{\mathcal{M}^{+}}e^{-iqx}+a^\dagger_{\mathcal{M}^{-}}e^{iqx}),\nonumber\\&&
\mathcal{M}_2(x)=\int{d^4{q}\over(2\pi)^4\sqrt{2m_{\mathcal{M}^{0}}}}
(a_{\mathcal{M}^{0}}e^{-iqx}+a^\dagger_{\bar
\mathcal{M}^{0}}e^{iqx}),
\end{eqnarray}
where $\mathcal{B}$ represents $\Lambda_c$ or $\Sigma_c$ and
$\mathcal{M}$ denotes $D$.

\subsection{The Bethe-Salpeter (B-S) equation for $\frac{1}{2}^-$ molecular state}
In the effective theory a meson and a baryon can interact via
exchanging hadrons. The Feynman diagram at the leading order is
depicted in Fig. \ref{penta1} (It is noted, the diagram where the
exchanged hadron is a heavy baryon, is ignored at the leading
order). The relative and total momenta of the bound state in the
equations are defined as
\begin{eqnarray} p = \eta_2p_1 -
\eta_1p_2\,,\quad q = \eta_2q_1 - \eta_1q_2\,,\quad P = p_1 + p_2
=q_1 + q_2 \,, \label{momentum-transform1}
\end{eqnarray}
where $p$ and $q$ are the relative momenta at the two sides of the
effective vertex, $p_1$ ($q_1$) and $p_2$ ($q_2$) are those
momenta of the constituents, $P$ is the total momentum of the
bound state, $k$ is the momentum of the exchanged meson, $\eta_i =
m_i/(m_1+m_2)$ and $m_i\, (i=1,2)$ is the mass of the $i$th
constituent meson.

The bound state composed of a baryon and a meson can be written as
\begin{eqnarray} \label{4-dim-BS1}
\chi_P(x_1,x_2)
=\langle0|T\mathcal{B}(x_1)\mathcal{M}(x_2)|P\rangle.
\end{eqnarray}

The B-S wave function is a Fourier transformation of that in momentum
space
\begin{eqnarray} \label{4-dim-BS3}
\chi_P(x_1,x_2) =e^{-iPX}\int\frac{d^4q}{(2\pi)^4}\chi_P(p).
\end{eqnarray}

By the so-called ladder approximation the corresponding B-S
equation was deduced in earlier references as
\begin{eqnarray} \label{4-dim-BS4}
\chi_{{P}}({ p})
=S_B(p_1)\int{d^4{q}\over(2\pi)^4}\,K(P,p,q)\chi_{_{P}}(q)S_M(p_2)\,,
\end{eqnarray}
where $S_B(p_1)$ is the propagator of the baryon ($\Lambda_c$ or
$\Sigma_c$), $S_M(p_2)$ is that of the meson ( $\bar D$) and
$K(P,p,q)$ is the kernel which can be obtained by calculating the
Feynman diagram in Fig. 1. For later convenience the relative
momentum $p$ is decomposed into the longitudinal $p_l$ ($\equiv
p\cdot v$) and transverse projection $p^\mu_t$ ($\equiv
p^\mu-p_lv^\mu$)=(0, $\mathbf{p}_T$) according to  the momentum of
the bound state $P$ ($v=\frac{P}{M}$).

\begin{eqnarray}\label{propagator1}
S_B(\eta_1
P+p)=\frac{i[(\eta_1M+p_l)v\!\!\!\slash+p_t\!\!\!\slash+m_1]}{(\eta_1M+p_l+\omega_l-i\epsilon)(\eta_1M+p_l-\omega_l+i\epsilon)},
\end{eqnarray}
\begin{eqnarray}\label{propagator2}
S_M(\eta_2
P-p)=\frac{i}{(\eta_2M-p_l+\omega_2-i\epsilon)(\eta_2M-p_l-\omega_2+i\epsilon)},
\end{eqnarray}
where $M$ is the total energy of the bound state, $\omega_i =
\sqrt{{ p_l}^2 + m_i^2}$ and $m_1$ ($m_2$) is the mass of the
baryon (meson).

By the Feynman diagram the kernel $K(P,p,q)$ is
written as
\begin{eqnarray}\label{kernel}
K(P,p,q)=-C_{I,I_z}g_{\mathcal{MMV}}g_{\mathcal{BBV}}(\gamma^\alpha-
\frac{\kappa_{\mathcal{BB}\rho}}{2m_\mathcal{B}}\sigma^{\alpha\beta}k_\beta)
(p_2+q_2)^\mu\Delta_{\alpha\mu}(k,m_V)F^2(k),
\end{eqnarray}
where $m_V$ is the mass of the exchanged meson, $g_{MMV}$,
$g_{BBV}$ and $\kappa_{BB\rho}$ are the concerned coupling
constants, $C_{I,I_z}$ is the isospin coefficient which is given
in Appendix B and
$\Delta_{\alpha\mu}(k,m_V)=(-g_{\alpha\mu}+k_{\alpha}k_{\mu}/m_v^2)/(k^2-m_v^2)$.
Apparently the contribution of the tensor term is much smaller
than that of the first term, thus we can ignore it in practical
computations. Indeed, a numerical estimate verifies this
allegation.

Since the constituents of the molecule (meson and baryon) are not
point particles,  a form factor at each effective vertex should be
introduced. The form factor suggested by many researchers is of
the form:
\begin{eqnarray} \label{form-factor} F({\bf k},m_{\rm V}^2 ) = {\Lambda^2 -
m_{\rm V}^2 \over \Lambda^2 + {\bf k}^2}\,,\quad {\bf k} = {\bf
p}-{\bf p}' \,,
\end{eqnarray}
where $\Lambda$ is a cutoff parameter. Since the form factor is
not derived from a fundamental principle, the concerned cutoff
parameter is neither determined theoretically, thus until now we
know little about the cutoff parameter $\Lambda$. In some
Refs.\cite{Meng:2007tk,Cheng:2004ru,Liu:2006df,Ke:2010aw} the form
factor is parameterized as $\lambda\Lambda_{QCD}+m_V$ with
$\Lambda_{QCD}=220$ MeV and the dimensionless parameter $\lambda$
is of order of unit. We will employ the expression
$\Lambda=\lambda\Lambda_{QCD}+m_V$ in our calculation.

The three-dimensional B-S wave function is obtained after
integrating over $p_l$
\begin{eqnarray} \label{3-dim-BS1}
\chi_{{P}}({ p_t}) =\int\frac{dp_l}{2\pi}\chi_{{P}}({ p}).
\end{eqnarray}

For the $S$wave system, the spatial wave function can be easily
derived \cite{Wang:2019krq,Weng:2010rb,Li:2019ekr}
\begin{eqnarray} \label{3-dim-BS2}
\chi_{{P}}({ p_t})
=[f_1(|\mathbf{p}_T|)+f_2(|\mathbf{p}_T|)p_t\!\!\!\slash]u(v,s),
\end{eqnarray}
where $f_1(|\mathbf{p}_T|)$ and $f_2(|\mathbf{p}_T|)$ are the
radial wave functions, $u(v,s)$, $v$ and $s$ are the spinor,
velocity and total spin of the pentaquark respectively.

Substituting Eq. (\ref{kernel}) into Eq. (9) and employing the
so-called covariant instantaneous approximation where $q_l=p_l$
i.e. $p_l$ takes the place of $q_l$ in the kernel $K(P,p,q)$,
$K(P,p,q)$  no longer depends on $q_1$. Then we are performing a
series of manipulations: integrate over $q_l$ on the right side of
Eq. (\ref{4-dim-BS4}); multiply $\int\frac{dp_l}{(2\pi)}$ on the
both sides of Eq. (\ref{4-dim-BS4}), and integrate over $p_l$ on
the left side using Eq. (\ref{4-dim-BS4}). Finally, substituting
Eq. (\ref{3-dim-BS2})  we obtain
\begin{eqnarray} \label{couple equation1}
&&[f_1(|\mathbf{p}_T|)+f_2(|\mathbf{p}_T|)p_t\!\!\!\slash]u(v,s)=-\int\frac{dp_l}
{(2\pi)}\int\frac{d^3\mathbf{q}_T}{(2\pi)^3}\frac{iC_{I,I_z}g_{MMV}g_{BBV}
[(\eta_1M+p_l)v\!\!\!\slash+p_t\!\!\!\slash+m_1]}
{[(\eta_1M+p_l)^2-\omega^2_l+i\epsilon][(\eta_1M-p_l)^2-\omega^2_2+i\epsilon)]}\nonumber\\&&\{-\frac{\kappa[2(\eta_2M-p_l)
v\!\!\!\slash-p_t\!\!\!\slash-q_t\!\!\!\slash](p_t\!\!\!\slash-q_t\!\!\!\slash)-(p_t\!\!\!\slash-q_t\!\!\!\slash)
[2(\eta_2M-p_l)v\!\!\!\slash-p_t\!\!\!\slash-q_t\!\!\!\slash]}{4m_B[-(\mathbf{p}_T-\mathbf{q}_T)^2-m_V^2]}+
\nonumber\\&&\frac{2(\eta_2M-p_l)v\!\!\!\slash-p_t\!\!\!\slash-q_t\!\!\!\slash-
(p_t\!\!\!\slash-q_t\!\!\!\slash)(\mathbf{p}_T^2-\mathbf{q}_T^2)/m_V^2}{-(\mathbf{p}_T-\mathbf{q}_T)^2-m_V^2}\}
F^2(k,m_V)[f_1(|\mathbf{q}_T|)+f_2(|\mathbf{q}_T|)q_t\!\!\!\slash]u(v,s).
\end{eqnarray}

Now let us finally fix the expressions of $f_1(|\mathbf{p}_T|)$
and $f_2(|\mathbf{p}_T|)$. Multiplying $\bar u(v,s)$ on both sides
of Eq.(\ref{couple equation1}), we get an expression which only
contains $f_1$ whereas multiplying $\bar u(v,s)p_t\!\!\!\slash$ to
the expression, $f_2$ is obtained, then by taking a trace, the
resultant formulas are
\begin{eqnarray} \label{couple equation1p}
&&f_1(|\mathbf{p}_T|)=-\int\frac{dp_l}{(2\pi)}\int\frac{d^3\mathbf{q}_T}{(2\pi)^3}\frac{iC_{I,I_z}g_{MMV}g_{BBV}
F^2(k,m_V)}
{[(\eta_1M+p_l)^2-\omega^2_l+i\epsilon][(\eta_1M-p_l)^2-\omega^2_2+i\epsilon]}\nonumber\\&&\{\frac{{\mathbf{p}_T\cdot
\mathbf{q}_T} + {\mathbf{p}_T}^2 + 2( {m_1} + {p_l} + M{\eta_1}) (
M{\eta_2}-{p_l}
 ) +{({\mathbf{p}_T}^2 -{\mathbf{p}_T\cdot \mathbf{q}_T}  )
( {\mathbf{p}_T}^2 -{\mathbf{q}_T}^2) }
 / {{m_V}^2}}{-(\mathbf{p}_T-\mathbf{q}_T)^2-m_V^2}f_1(|\mathbf{q}_T|)+\nonumber\\&&\frac{\frac{( {\mathbf{p}_T}^2 -{\mathbf{q}_T}^2)
   (\mathbf{p}_T\cdot \mathbf{q}_T - {\mathbf{q}_T}^2 )
    ( {m_1} + {p_l} + M{\eta_1} ) }{{m_V}^2}
  + (m_1+M\eta_1+p_l)(
{\mathbf{p}_T\cdot \mathbf{q}_T} + {\mathbf{q}_T}^2 )  +
2(M{\eta_2}-p_l)\mathbf{p}_T\cdot
\mathbf{q}_T}{-(\mathbf{p}_T-\mathbf{q}_T)^2-m_V^2}f_2(|\mathbf{q}_T|)\nonumber\\&&+\frac{\kappa}{m_B}\frac{[
{{\mathbf{p}_T\cdot \mathbf{q}_T}}^2 -
{\mathbf{p}_T}^2{\mathbf{q}_T}^2 + (m_1+p_l+M\eta_1)(
{\mathbf{q}_T}^2 -{\mathbf{p}_T\cdot \mathbf{q}_T} )(p_l-M\eta_2)
    ]f_2(|\mathbf{q}_T|)}{[-(\mathbf{p}_T-\mathbf{q}_T)^2-m_V^2]}\nonumber\\&&-\frac{\kappa}{m_B}\frac{(
{\mathbf{p}_T\cdot \mathbf{q}_T} - {\mathbf{p}_T}^2 ) ( {p_l} -
M{\eta_2}
)f_1(|\mathbf{q}_T|)}{[-(\mathbf{p}_T-\mathbf{q}_T)^2-m_V^2]}\}.
\end{eqnarray}

\begin{eqnarray} \label{couple equation2}
&&f_2(|\mathbf{p}_T|){\mathbf{p}_T}^2=-\int\frac{dp_l}{(2\pi)}\int\frac{d\mathbf{q}_T}{(2\pi)^3}\frac{-iC_{I,I_z}g_{MMV}g_{BBV}
F^2(k,m_V)}
{[(\eta_1M+p_l)^2-\omega^2_l+i\epsilon][(\eta_1M-p_l)^2-\omega^2_2+i\epsilon]}\nonumber\\&&\{\frac{{-{\mathbf{p}_T}^2(\mathbf{p}_T\cdot
\mathbf{q}_T} + {\mathbf{q}_T}^2) + 2{\mathbf{p}_T\cdot
\mathbf{q}_T}( {m_1}-{p_l} - M{\eta_1}) ( M{\eta_2}-{p_l}
 ) +{\mathbf{p}_T}^2\frac{({\mathbf{q}_T}^2 -{\mathbf{p}_T\cdot \mathbf{q}_T}  )
( {\mathbf{p}_T}^2 -{\mathbf{q}_T}^2) }
  {{{m_V}}^2}}{-(\mathbf{p}_T-\mathbf{q}_T)^2-m_V^2}f_2(|\mathbf{q}_T|)+\nonumber\\&&\frac{\frac{( {\mathbf{p}_T}^2 -{\mathbf{q}_T}^2) (\mathbf{p}_T\cdot \mathbf{q}_T - {\mathbf{p}_T}^2 )
    ( -{m_1} + {p_l} + M{\eta_1} ) }{{{m_V}}^2}
  + (m_1-M\eta_1+p_l)(
{\mathbf{p}_T\cdot \mathbf{q}_T} + {\mathbf{p}_T}^2 )  -
2M{\eta_2}{\mathbf{p}_T}^2-2p_l\mathbf{p}_T\cdot
\mathbf{q}_T}{-(\mathbf{p}_T-\mathbf{q}_T)^2-m_V^2}f_1(|\mathbf{q}_T|)\nonumber\\&&-\frac{\kappa}{m_B}\frac{[(
{M\eta_2} - {pl} ) {\mathbf{p}_T\cdot
\mathbf{q}_T}{\mathbf{p}_T}^2 -
 {\mathbf{p}_T\cdot \mathbf{q}_T}^2( {m_1} - {p_l}- M{\eta_1} )  +
  {\mathbf{p}_T}^2{\mathbf{q}_T}^2( {m_1} - M
     )]f_2(|\mathbf{q}_T|)}{[-(\mathbf{p}_T-\mathbf{q}_T)^2-m_V^2]}\nonumber\\&&-\frac{\kappa}{m_B}\frac{(
{\mathbf{p}_T\cdot \mathbf{q}_T} - {\mathbf{p}_T}^2 ) ( {p_l} -
M{\eta_2}
)(m_1-p_l-M\eta_1)f_1(|\mathbf{q}_T|)}{[-(\mathbf{p}_T-\mathbf{q}_T)^2-m_V^2]}\}.
\end{eqnarray}
To extract $f_1(|\mathbf{p}_T|)$ and $f_2(|\mathbf{p}_T|)$ from the
above equations, instead of the procedure adopted in earlier works,
we multiply $\bar u(v)$ from the right side of the equation and sum
over the spin projections of $u(v)$, then taking a trace of the
modified equation, the job is done. The advantage of this procedure
is to keep the equation of motion $v\!\!\!\slash u(v,s)=u(v,s)$.

Now we perform an integral over $p_l$ on the right side of Eqs.
(\ref{couple equation1}) and (\ref{couple equation2}) where four
poles exist at $-\eta_1M-\omega_1+i\epsilon$,
$-\eta_1M+\omega_1-i\epsilon$, $\eta_2M+\omega_2-i\epsilon$ and
$\eta_2M-\omega_2+i\epsilon$. By choosing an appropriate contour
(\ref{couple equation1}) and (\ref{couple equation2}) we calculate
the residuals at $p_l=-\eta_1M-\omega_1+i\epsilon$ and
$p_l=\eta_2M-\omega_2+i\epsilon$. The coupled equations after the
contour integrations are collected in the appendix [Eqs.
(\ref{couple equation12}) and (\ref{couple equation22}) ]. Then
one can carry out the azimuthal integration and reduce Eqs.
(\ref{couple equation12}) and (\ref{couple equation22}) to
one-dimensional integral equations
\begin{eqnarray}\label{one dimension equation}
&&f_1(|\mathbf{p}_T|)=\int{d|\mathbf{q}_T|}[A_{11}(|\mathbf{q}_T|,|\mathbf{p}_T|)
f_1(|\mathbf{q}_T|)+A_{12}(|\mathbf{q}_T|,|\mathbf{p}_T|)f_2(|\mathbf{q}_T|)]\nonumber\\&&
f_2(|\mathbf{p}_T|)=\int{d|\mathbf{q}_T|}[A_{21}(|\mathbf{q}_T|,|\mathbf{p}_T|)
f_1(|\mathbf{q}_T|)+A_{22}(|\mathbf{q}_T|,|\mathbf{p}_T|)f_2(|\mathbf{q}_T|)],
\end{eqnarray}
where $A_{11}$, $A_{12}$, $A_{21}$ and $A_{22}$ are presented in
the appendix [see Eqs. (\ref{couple equation13}), (\ref{couple
equation14}), (\ref{couple equation23}) and (\ref{couple
equation24})].

\subsection{The normalization condition for the B-S wave function}
The normalization condition for the B-S wave function of a bound
state is\cite{Guo:2007mm,Weng:2010rb}
\begin{eqnarray}\label{normal1}
i\int \frac{d^4pd^4q}{(2\pi)^8}\bar
\chi_P(p)\frac{\partial}{\partial
P_0}[I(P,p,q)+K(P,p,q)]\chi_P(q)=1,
\end{eqnarray}
where $P_0$ is the energy of the bound state and the spinor relation
$\sum_s u(v,s)\bar u(v,s)=\frac{v\!\!\!\slash+1}{2}$ is used.
$I(P,p,q)$ is the reciprocal of the four-point propagator
\begin{eqnarray}
I(P,p,q)=\frac{\delta^4(p-q)}{(2\pi)^4}[S_\mathcal{B}(p_1)]^{-1}[S_\mathcal{M}(p_2)]^{-1}.
\end{eqnarray}

For the molecular sates composed of two mesons the second term in
the normalization condition is several orders smaller than the
first term \cite{Ke:2012gm,Ke:2018jql}; thus we have all every
reason to believe that the rule also applies to the case where the
molecule is composed of a baryon and a meson, consequently the
term $\frac{\partial}{\partial P_0}K(P,p,q)$  can be ignored and
then
\begin{eqnarray}\label{normal1}
-\int \frac{d^4p}{(2\pi)^4}\bar
\chi_P(p)\eta_1v\!\!\!\slash[S_\mathcal{M}(p_2)]^{-1}\chi_P(q)-\int
\frac{d^4p}{(2\pi)^4}\bar \chi_P(p)2\eta_2p_2\cdot
v[S_\mathcal{B}(p_1)]^{-1}\chi_P(q)=1.
\end{eqnarray}
Let us define the transverse projections of the B-S wave function
as follows:
\begin{eqnarray}\label{normal2}
&&\alpha_P(p)=-i[S_\mathcal{B}(p_1)]^{-1}\chi_P(q)[S_\mathcal{M}(p_2)]^{-1},\nonumber\\
&&\beta_P(p)=-i[S_\mathcal{M}(p_2)]^{-1}\bar\chi_P(q)[S_\mathcal{B}(p_1)]^{-1},
\end{eqnarray}
the normalization condition is
\begin{eqnarray}\label{normal3}
&&-\int \frac{d^4p}{(2\pi)^4}{\rm
Tr}[\alpha_P(p)\beta_P(p)S_\mathcal{B}(p_1)\eta_1v\!\!\!\slash
S_\mathcal{B}(p_1)S_\mathcal{M}(p_2)]\nonumber\\&&-\int
\frac{d^4p}{(2\pi)^4}{\rm Tr}[\alpha_P(p)\beta_P(p)2\eta_2p_2\cdot
v S_\mathcal{B}(p_1) S_\mathcal{M}(p_1) S_\mathcal{M}(p_1)]=1.
\end{eqnarray}

Substituting the expression $\bar \chi_P(p)$ [Eq.
(\ref{4-dim-BS4})] into Eqs. (\ref{normal2}) under the covariant
instantaneous approximation one can obtain the expressions of
$\alpha_P(p)$ and $\beta_P(p)$, for example
\begin{eqnarray} \label{alpha equation1}
&&\alpha_P(p)=-\int\frac{d^3\mathbf{q}_T}{(2\pi)^3}{C_{I,I_z}g_{\mathcal{MMV}}
g_{\mathcal{BBV}}\{\frac{2(\eta_2M-p_l)v\!\!\!\slash-p_t\!\!\!\slash-q_t
\!\!\!\slash-(p_t\!\!\!\slash-q_t\!\!\!\slash)
(\mathbf{p}_T^2-\mathbf{q}_T^2)/m_V^2}{-(\mathbf{p}_T-\mathbf{q}_T)^2-m_V^2}
} \nonumber\\&&\-\frac{\kappa[2(\eta_2M-p_l)
v\!\!\!\slash-p_t\!\!\!\slash-q_t\!\!\!\slash](p_t\!\!\!\slash-q_t\!\!\!\slash)-(p_t\!\!\!\slash-q_t\!\!\!\slash)
[2(\eta_2M-p_l)v\!\!\!\slash-p_t\!\!\!\slash-q_t\!\!\!\slash]}{4m_B[-(\mathbf{p}_T-\mathbf{q}_T)^2-m_V^2]}
\nonumber\\&&\}
F^2(k,m_V)[f_1(|\mathbf{q}_T|)+f_2(|\mathbf{q}_T|)q_t\!\!\!\slash]u(v,s).
\end{eqnarray}
and $\alpha_P(p)$ and $\beta_P(p)$ can be parameterized into
\begin{eqnarray}\label{normal21}
&&\alpha_P(p)=[h_1(|\mathbf{p}_T|)+h_2(|\mathbf{p}_T|)p_t\!\!\!\slash]u(v,s),\nonumber\\&&\beta_P(p)=\bar
u(v,s)[h_1(|\mathbf{p}_T|)+h_2(|\mathbf{p}_T|)p_t\!\!\!\slash],
\end{eqnarray}
with
\begin{eqnarray}\label{normal22}
&&h_1(|\mathbf{p}_T|)=-\int\frac{d^3\mathbf{q}_T}{(2\pi)^3}\{\frac{{
2{f_1(\mathbf{q}_T)}(M{\eta_2}-{p_l} )+f_2(|\mathbf{q}_T|)}[
{\mathbf{q}_T^2}+\mathbf{p}_T\cdot\mathbf{q}_T + \frac{(
{\mathbf{p}_T^2} - {\mathbf{q}_T^2})
(\mathbf{p}_T\cdot\mathbf{q}_T -{\mathbf{q}_T^2} ) }
      {{{m_V}}^2} ]
}{-(\mathbf{p}_T-\mathbf{q}_T)^2-m_V^2}\nonumber\\&&+\frac{4\kappa[f_2(|\mathbf{q}_T|)(
(M{\eta_2}-p_l)(\mathbf{p}_T\cdot \mathbf{q}_T- {\mathbf{q}_T^2}
)]}{4m_B[-(\mathbf{p}_T-\mathbf{q}_T)^2-m_V^2]}\}{C_{I,I_z}g_{\mathcal{MMV}}g_{\mathcal{BBV}}
}F^2(k,m_V),\nonumber\\&&h_2(|\mathbf{p}_T|)=-\int\frac{d^3\mathbf{q}_T}{(2\pi)^3}\{\frac{f_1(\mathbf{q}_T)(
\frac{(\mathbf{p}_T\cdot\mathbf{q}_T -{\mathbf{p}_T^2} )
        (\mathbf{p}_T^2 - \mathbf{q}_T^2 ) }{{{m_V}}^2}-\mathbf{p}_T\cdot\mathbf{q}_T -{\mathbf{p}_T^2} )  +
2f_2(\mathbf{q}_T)( {p_l} -
M{\eta_2})\mathbf{p}_T\cdot\mathbf{q}_T}{\mathbf{p}_T^2[-(\mathbf{p}_T-\mathbf{q}_T)^2-m_V^2]}\nonumber\\&&+\frac{4\kappa(
[f_2(\mathbf{q}_T)(
{\mathbf{p}_T^2}{\mathbf{q}_T^2}-{\mathbf{p}_T\cdot\mathbf{q}_T}^2
)  +
  f_1(\mathbf{q}_T)(\mathbf{p}_T\cdot\mathbf{q}_T -{\mathbf{p}_T^2} ) ({p_l} - M{\eta_2} )
]}{4m_B\mathbf{p}_T^2[-(\mathbf{p}_T-\mathbf{q}_T)^2-m_V^2]}\}\nonumber\\&&{C_{I,I_z}g_{\mathcal{MMV}}g_{\mathcal{BBV}}
F^2(k,m_V)},
\end{eqnarray}
Substituting Eqs. (\ref{propagator1}), (\ref{propagator2}) and
equation group (\ref{normal21}) into Eq. (\ref{normal3}) we
obtain
\begin{eqnarray}\label{normal4}
&&i\int \frac{d^4p}{(2\pi)^4}2\{
    {{h_1}}^2[ {{m_1}}^2 + {{p_l}}^2 + {\mathbf{p}_T^2} + 2M{p_l}{\eta_1} + M^2{{\eta_1}}^2 +
       2{m_1}( {p_l} + M{\eta_1} )  ]+\nonumber\\&&
    {{h_2}}^2{\mathbf{p}_T^2}( {{m_1}}^2 + {{p_l}}^2 + {\mathbf{p}_T^2} + 2M{p_l}{\eta_1} +
       M^2{{\eta_1}}^2 - 2{m_1}( {p_l} + M{\eta_1} )  )  -4{h_1}{h_2}{\mathbf{p}_T^2}( {p_l}
+ M{\eta_1} )
\}\nonumber\\&&/\{[(\eta_1M+p_l)^2-\omega^2_l+i\epsilon]^2[(\lambda_1M-p_l)^2-\omega^2_2+i\epsilon)]\}\nonumber\\&&+i\int
\frac{d^4p}{(2\pi)^4}2[
 {{h_1}}^2( {m_1} + {p_l} + M{\eta_1}
)-2{h_1}{h_2}{\mathbf{p}_T^2} + {{h_2}}^2{\mathbf{p}_T^2}
     (  {p_l}-{m_1}  + M{\eta_1} )   ]\nonumber\\&&/\{[(\eta_1M+p_l)^2-\omega^2_l+
i\epsilon]^2[(\lambda_1M-p_l)^2-\omega^2_2+i\epsilon)]\}=1.
\end{eqnarray}
After the contour integration on $p_l$ and the azimuthal
integration the normalization condition can be calculated
numerically and the values of $f_1(|\mathbf{p}_T|)$ and
$f_2(|\mathbf{p}_T|)$ are fixed at the same time.

\subsection{the decay of $P_c\to \mathcal{V}$+proton}

Now we investigate the strong decays of $P_c$ in terms of the
framework formulated above.

\begin{figure*}
        \centering
        \subfigure[~]{
          \includegraphics[width=7cm]{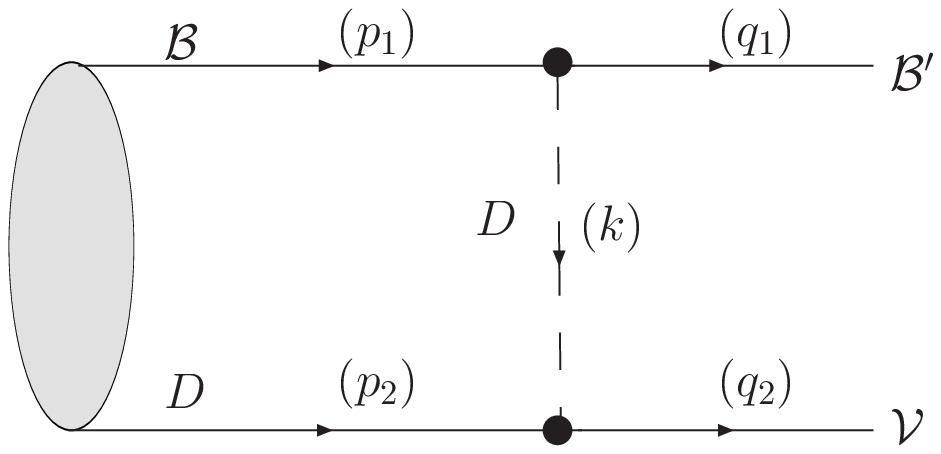}}
        \subfigure[~]{
          \includegraphics[width=7cm]{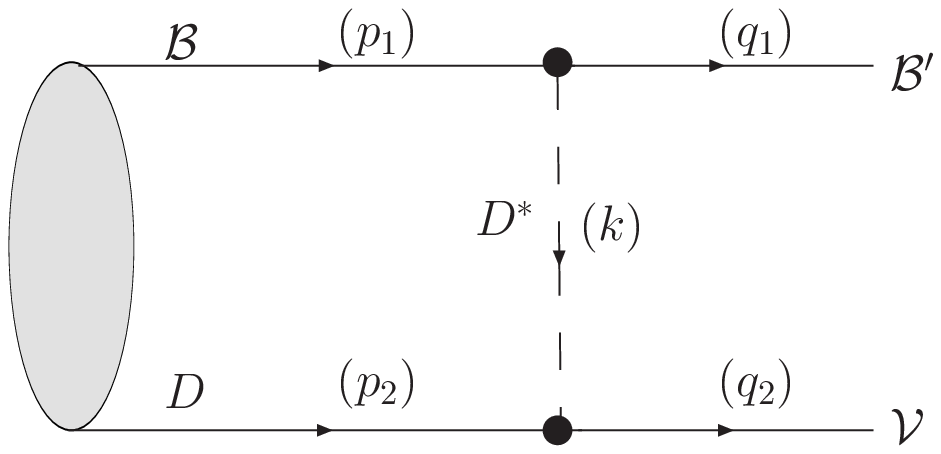}}
 \caption{the decay of  $P_c$  by exchanging  mesons .}
        \label{decay}
    \end{figure*}

The amplitudes corresponding to the two diagrams in Fig. \ref{decay}
are,
\begin{eqnarray}\label{a1}
\mathcal{A}_{a}=C_I{g_{\mathcal{B}\mathcal{B}'
D}g_{DD\mathcal{V}}}\int\frac{d^4p}{(2\pi)^4}\bar
U_{\mathcal{B}'}\gamma^5
\chi_P(p)(k-p_2)_\nu{\epsilon}^\nu\frac{1}{k^2-M^2_D}F^2(k,m_D),
\end{eqnarray}
\begin{eqnarray}\label{a2}
\mathcal{A}_{b}&&=2C_I{g_{\mathcal{B}\mathcal{B}'
D^*}g_{DD^*\mathcal{V}}}\int\frac{d^4p}{(2\pi)^4}\bar
U_{\mathcal{B}'}(\gamma^\sigma-\frac{\kappa_{BBD^*}}{2m_B}\sigma^{\sigma\omega}k_\omega)
\chi_P(p)\varepsilon^{\alpha\beta\mu\nu}k_{\mu}{q_2}_\alpha{\epsilon_{\nu}}\\&&\frac{g_{\sigma\beta}-k_\beta
k_\sigma/M_{D^*}^2}{k^2-M^2_{D^*}}F^2(k,m_{D^*}),
\end{eqnarray}
where $C_I$ is the isospin coefficient of the transition,
$k=p-(\eta_2 q_1-\eta_1 q_2)$, $\mathcal{B}$ denotes the charmed baryon in the molecular
state: $\Sigma_c$ or $\Lambda_c$; $\epsilon$ is the polarization
vector of $\mathcal{V}$ and $\mathcal{B'}$ represents proton. We
still take the approximation $k_0=0$ to carry out the calculation.

The total amplitude is
\begin{eqnarray}\label{a3}
\mathcal{A}=\mathcal{A}_a+\mathcal{A}_b= \bar
u_{\mathcal{B}'}[\gamma^5g_1\gamma^\mu+i\gamma^5g_2\sigma^{\mu\nu}
q_{2\nu}+ig_3\gamma_\nu\varepsilon^{\mu\nu\alpha\beta}P_\alpha
q_{2\beta}] u(v) {\epsilon_{\mu}}.
\end{eqnarray}
The factors $g_1$, $g_2$ and $g_3$ can be extracted from the
expressions of $\mathcal{A}_1$ and $\mathcal{A}_2$.

 Then the partial width is expressed as
\begin{eqnarray}
d\Gamma=\frac{1}{32\pi^2}|\mathcal{A}|^2\frac{|q_2|}{M^2}d\Omega.
\end{eqnarray}

\section{numerical results}
\subsection{the numerical results }

In order to solve the B-S equation numerically some parameters are
needed. The mass $m_{\Lambda_c}$, $m_{\Sigma_c}$, $m_D$,
$m_\omega$, $m_\rho$ is taken from the databook\cite{PDG18}.
Following Ref.\cite{Ronchen:2012eg,Shen:2017ayv}, we set the
coupling constants $g_{DD\omega}=g_{DD\rho}=3.02$,
$g_{\Lambda_c\Lambda_c\omega}=8.125$,
$g_{\Sigma_c\Sigma_c\omega}=g_{\Sigma_c\Sigma_c\rho}=7.475$,
$f_{\Sigma_c\Sigma_c\omega}=\kappa
g_{\Sigma_c\Sigma_c\omega}=9.9125$,
$f_{\Sigma_c\Sigma_c\rho}=\kappa g_{\Sigma_c\Sigma_c\rho}=9.9125$
.

With these parameters and the corresponding isospin factors a
complete B-S equation [the coupled equations (\ref{one dimension
equation})] is established. These coupled equations are
complicated integral equations, thus to numerically solve them,
the standard way is to discretize them, namely we would convert
them into algebraic  equations. Concretely,  we set a reasonable
finite range for $\bf |p_T|$ and $\bf |q_T|$, and let the
variables take $n$ ( $n$=129 in our calculation) discrete values
$Q_1$, $Q_2$,...$Q_n$ which distribute with equal gap from
$Q_1$=0.001 GeV to $Q_n$=2 GeV . The gap between two adjacent
values is $\Delta \bf |p_T|$=(1.999/128) GeV. For clarity, we let
$n$ values of $f_1({\bf |p_T|})$ and $n$ values of $f_2({\bf
|p_T|})$ constitute a column matrix with $2n$ rows and the $2n$
elements $f_1({\bf |q_T|})$, $f_2({\bf |q_T|})$ construct another
column matrix residing on the right side of the equation as shown
below. The column matrix composed of $f_1({\bf |p_T|})$ and
$f_2({\bf |p_T|})$ is associated with the right column matrix of
$f_1({\bf |q_T|})$ and $f_2({\bf |q_T|})$ by a $2n\times 2n$
matrix whose elements are the coefficients given in Eq. (\ref{one
dimension equation}). The standard way to treat the equation is to
let $\bf |p_T|$ and $\bf |q_T|$ take the same sequential values
$Q_1$, $Q_2$,...$Q_n$ for discretizing the integral equation.
$$\left(\begin{array}{c}
      f_1(Q_1) \\... \\f_1(Q_{129})\\
        f_2(Q_1)\\... \\f_2(Q_{129})
      \end{array}\right)=A(\Delta E, \lambda)\left(\begin{array}{c}
     f_1(Q_1) \\... \\f_1(Q_{129})\\
        f_2(Q_1)\\... \\f_2(Q_{129})
      \end{array}\right).$$
As a matter of fact, it is a homogeneous linear equation group. If
it possesses non-trivial solutions, the necessary and sufficient
condition is the coefficient determinant to be zero. In our case,
it is $|A(\Delta E,\lambda)-I|=0$ ($I$ is the unit matrix) where
$A(\Delta E,\lambda)$. Now we calculate the determinant of
$|A(\Delta E,\lambda)-I|$  is a function of the binding energy
$\Delta E=m_1+m_2-M$ and parameter $\lambda$. Our strategy is
following: we arbitrarily vary $\Delta E$ within a possible range,
by requiring $|A(\Delta E,\lambda)-I|=0$, we obtain a
corresponding $\lambda$. In Ref.\cite{Meng:2007tk} $\lambda$ was
fixed to be 3. In our earlier paper\cite{Ke:2010aw} we change the
value of $\lambda$ from 1 to 3 to explore possible dependence of
the results on it,  it seems that a value of $\lambda$ within the
range of $0\sim 4$ is reasonable for forming a bound state of two
hadrons. Consequently, if the obtained $\lambda$ is much beyond
the range, we would conclude that the resonance cannot exist.

To get the wavefunction
$T(f_1(Q_1),f_1(Q-2)...,f_2(Q_1)...f_2(Q_{129})$, we adopt a
special method. Namely, we suppose a matrix equation $(A(\Delta
E,\lambda)_{ij})(f(j))=\beta (f(i))$ which just is an eigenequation. In
terms of the standard software, we can find all the possible
``eigenvalues" $\beta$, and among them only $\beta=1$ is the solution
we expect, then the corresponding wavefunction is gained which
just is the solution of the B-S equation.

For $|A(\Delta E,\lambda)-I|=0$, inputting some  binding energies,
we would check whether we can obtain reasonable values for
$\lambda$. If yes, we substitute the values of $\lambda$ and the
binding energy into the matrix equation to obtain the B-S
wavefunctions. With this strategy, we investigate the molecular
structure of $\Lambda_c$ and $\bar D$ as well as that of
$\Sigma_c$ and $\bar D$.

If the exchanging particles are limited to light vector meson,
only $\omega$ and $\rho$ can be exchanged between charmed baryons
and $D$. Of course, exchanging two $\rho$ mesons between
$\Lambda_c$ ($\Sigma_c$) and $\bar D$ can also induce a potential,
but it undergoes a loop  suppression, therefore, we do not
consider that contribution.

As the first trial, let us study a simple compound, namely
we explore the possible bound states
of $\Lambda_c$ and $\bar D$ which is an $I=\frac{1}{2}$ state, therefore
only $\omega$ can be exchanged between $\Lambda_c$ and $\bar D$.
We find that there is no solution for the B-S equation, therefore we
would conclude that the interaction induced by the single $\omega$ exchange
is repulsive.

With the same procedure, we study a molecule composed of
$\Sigma_c$ and $\bar D$ whose isospin could be either $1/2$ or
$3/2$ and the coefficient is $C_{\frac{1}{2},\frac{1}{2}}=1$.
Since $P_C(4312)$ is observed in the $J/\psi p$ portal, it is
confirmed to be a state of $I=1/2$. In this case both $\omega$ and
and $\rho$ exchanges between the two ingredients are allowed. The
isospin factor for the $\rho$ exchange is $-2$, namely plays an
opposite role to the $\omega$ exchange. We try to solve the
equation $|A(\Delta E,\Lambda)-I|=0$ for some chosen $\Delta E$
and find a solution for $\Sigma_c \bar D$ with the quantum number
$I(J)={1\over 2}({1\over 2})$ where the factor $\lambda$ can span
a large range.

The result indicates that although, the $\omega$ exchange
contributes a repulsive interaction, for $\Sigma_c \bar D$
molecule, the total interaction can be attractive due to a larger
contribution from the $\rho$ exchange. Numerically, the obtained
values of $\lambda$ and corresponding $\Delta E$ for $\Sigma_c\bar
D$ system are presented in Table \ref{Tab:12}. Our numerical
computation also confirms that the tensor coupling in the
$\mathcal{L}_{\mathcal{BBV}}$ has little effect on the results.
For example setting $\Delta E=8$ MeV one can fix $\lambda=3.77$
and 3.88  with and without the tensor contribution the obtained
wave functions are very close to each other so we can safely
ignore the tensor coupling in the vertex
$\mathcal{L}_{\mathcal{BBV}}$. Apparently when $\Delta E$ is very
small the obtained $\lambda$ is smaller than 4, so $\Sigma_c$ and
$\bar D$ should form a weak bound state. At present  the
pentaquark $P_c(4312)$ has been experimentally observed in
$\Lambda_b\to J/\psi pK$ portal, which is peaked at the invariant
mass spectrum of $J/\psi p$ and has the invariant mass of about
4312 MeV. Apparently its isospin is $\frac{1}{2}$, and the
majority of authors
\cite{Chen:2019bip,He:2019ify,Xiao:2019mst,Zhang:2019xtu,Wang:2019hyc,Xu:2019zme,Wang:2019got}
regarded this pentaquark as a bound state of $\Sigma_c$ and $\bar
D$ and we agree with it.

Using the normalized wave functions the transition
$P_{c(\frac{1}{2},\frac{1}{2})}\to J/\psi+p$ is calculable. The
form factors defined in Eq. (\ref{a3}) with the coupling constants
are evaluated: $g_{\mathcal{B}\mathcal{B}'D}=2.7$,
$g_{\mathcal{B}\mathcal{B}'D^*}=3.0$, $g_{DD\psi}=7.4$,
$g_{DD^*\psi}=2.5$ GeV$^{-1}$\cite{Shen:2016tzq}. We obtain
$g_1=0.396$ GeV, $g_2=0.270$, $g_3=0.00632$ GeV$^{-1}$ and the
decay width $\Gamma[P_{c(\frac{1}{2},\frac{1}{2})}\to J/\psi
p]=3.66$ MeV. If the binding energy is 20 MeV, $g_1=0.412$ GeV,
$g_2=0.282$, $g_3=0.00923$ GeV$^{-1}$ can be obtained and the
estimated decay width is $\Gamma[P_{c(\frac{1}{2},\frac{1}{2})}\to J/\psi
p]=3.90$ MeV.  We notice that our results are close to that of
Ref.\cite{Xiao:2019mst,Xu:2019zme}, but the results given in
Ref.\cite{Lin:2019qiv} are 1-3 orders smaller than ours where
different ultraviolet regulators are employed.

By our observation given above, for the state with $I=\frac{3}{2}$
the isospin factor is 1 for  exchanging
either  $\omega$ or
$\rho$, therefore the total interaction is repulsive, it means that
$\Sigma_c$ and $\bar D$ cannot form a bound state
with $I=\frac{3}{2}$.

\begin{table}
\caption{the cutoff parameter $\lambda$ and the corresponding
binding energy $\Delta E$ for the bound state $\Sigma_c \bar D$
with $I=\frac{1}{2}$ and $I_z=\frac{1}{2}$}\label{Tab:12}
\begin{ruledtabular}
\begin{tabular}{cccccccc}
  $\Delta E$ MeV  &2 & 8 &  20  &  30& 40 \\\hline
  $\lambda$  & 3.31 & 3.88    &4.58 &5.04   &5.44
\end{tabular}
\end{ruledtabular}
\end{table}
\begin{figure}[hhh]
\begin{center}
\scalebox{0.8}{\includegraphics{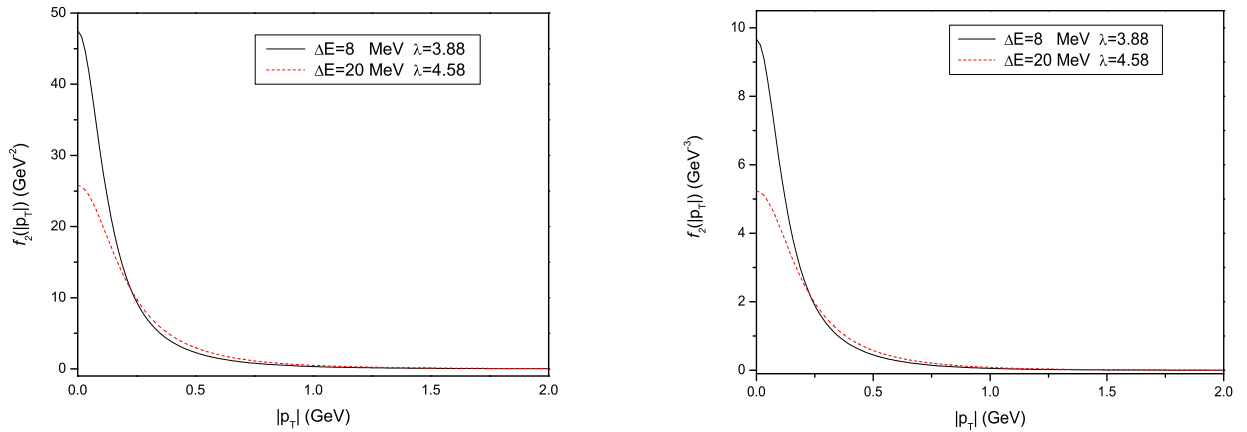}}
\end{center}
\caption{The normalized wave function $f_1(|\mathbf{p}_T|)$ and
$f_2(|\mathbf{p}_T|)$ for $P_{c(\frac{1}{2},\frac{1}{2})}$
}\label{wave}
\end{figure}

\subsection{predictions about pentaquark $P_b$}
The isospin of the $\Lambda_b B^+$ system is
\begin{eqnarray}  \label{isospin11}
|\frac{1}{2},\frac{1}{2}\rangle=|\Lambda_b B^+\rangle.
\end{eqnarray}

The isospin of the $\Sigma_b B$ system can be
$|\frac{1}{2},\pm\frac{1}{2}\rangle$
 $|\frac{3}{2},\pm\frac{1}{2}\rangle$ and $|\frac{3}{2},\pm\frac{3}{2}\rangle$. Let us work on the isospin states
\begin{eqnarray}  \label{isospin21}
|\frac{1}{2},\frac{1}{2}\rangle=\sqrt{\frac{2}{3}}|\Sigma_b^{+}
B^0\rangle-\sqrt{\frac{1}{3}}|\Sigma_b^0 B^+\rangle,
\end{eqnarray}
\begin{eqnarray}  \label{isospin31}
|\frac{3}{2},\frac{1}{2}\rangle=\sqrt{\frac{1}{3}}|\Sigma_b^{+}
B^0\rangle+\sqrt{\frac{2}{3}}|\Sigma_b^0 B^+\rangle,
\end{eqnarray}
and
\begin{eqnarray}  \label{isospin41}
|\frac{3}{2},\frac{3}{2}\rangle=|\Sigma_b^{+} B^+\rangle.
\end{eqnarray}

\begin{table}
\caption{the cutoff parameter $\lambda$ and the corresponding
binding energy $\Delta E$ for the bound state $\Sigma_b  B$ with
$I=\frac{1}{2}$ and $I_z=\frac{1}{2}$}\label{Tab:t21}
\begin{ruledtabular}
\begin{tabular}{cccccccc}
  $\Delta E$   & 10 &  20  &  30& 40 &50\\\hline
  $\lambda$  & 2.13   &2.51 &2.82   &3.09 &3.35
\end{tabular}
\end{ruledtabular}
\end{table}

Using the masses of $\Lambda_b$, $\Sigma_b$ and $B$ presented in
Ref.\cite{PDG18} and other parameters listed in previous sections,
we solve those B-S equations. It is found that only the equation
for the $\Sigma_b  B$ system with $I=\frac{1}{2}$ has a solution.
The binding energy and corresponding $\lambda$ values are
presented in Table \ref{Tab:t21}. That implies that the bound
state with $I=\frac{1}{2}$ can exist in the nature. Under the
heavy quark symmetry we suppose that the couplings are unchanged
when $b$-hadrons replace $c$-hadrons. We turn to study the
transition $P_{b(\frac{1}{2},\frac{1}{2})}\to \Upsilon p$. We
obtain $g_1=0.00346$ GeV, $g_2=0.252$, $g_3=0.0000911$ GeV$^{-1}$
and predict the decay width
$\Gamma[P_{b(\frac{1}{2},\frac{1}{2})}\to \Upsilon p]=0.690$ keV
as the binding energy is 10 MeV. If $\Delta E=20$ MeV the decay
width $\Gamma[P_{b(\frac{1}{2},\frac{1}{2})}\to \Upsilon p]=1.09$
keV and $g_1=0.00435$ GeV, $g_2=0.318$, $g_3=0.000149$ GeV$^{-1}$.

\section{conclusion and discussion}

Within the B-S framework we explore several bound states which are
composed of a baryon and a meson. Their total spin and parity is
$\frac{1}{2}^-$ i.e. the orbital angular momentum $L=0$
($S$-wave). We try to solve  the B-S equation for getting possible
spatial wave functions for $\Lambda_c \bar D$, $\Sigma_c \bar D$,
$\Lambda_b  B$ and $\Sigma_b  B$ systems. If the B-S equation for
a supposed molecular structure has a stable solution, we would
conclude that the concerned pentaquark could exist in the nature,
oppositely, no-solution means the supposed pentaquark cannot
appear as a resonance or the molecular state is not an appropriate
structure. The solution can apply as a criterion for the
structures of the pentaquark states which have already been or
will be experimentally observed. In this scenario, the two
constituents interact by exchanging light vector mesons. For the
$\Lambda_c\bar D$ ($\Lambda_b B$) system only $\omega$ is the
exchanged mediate meson, while for the $\Sigma_c\bar D$ system
($\Sigma_b B$) both $\omega$ and $\rho$ contribute. The chiral
interaction determines if those molecular states can be formed.

For  $\frac{1}{2}^-$ baryon ($S$-wave), the B-S wave function
possesses two scalar functions $f_1(|\mathbf{p_T}|)$ and
$f_2(|\mathbf{p_T}|)$ which should be solved numerically.
Discretizing the integral equations, we simplify the B-S equation
into two coupled algebraic equations about $f_1(|\mathbf{p_T}|)$
and $f_2(|\mathbf{p_T}|)$.

As $|\mathbf{p_T}|\;\; (i=1,2)$ takes $n$ discrete values the two
coupled equations are converted into a matrix equation which can
be easily solved numerically in terms of available softwares. When
all known parameters are input there still is one undetermined
parameter $\lambda$. Our strategy is inputting binding energies
within a range and then fixing $\lambda$ by solving the matrix
equation. If $\lambda$ is located in a reasonable range one can
expect the bound state to exist. We find the B-S equation of  the
state $\Lambda_c \bar D$ system has no solution for $\lambda$ when
the binding energy takes experimentally allowed values. For the
$\Sigma_c \bar D$ system there are three isospin eigenstates. Due
to the isospin factors, the B-S equations for
$P_{c(\frac{1}{2},\frac{1}{2})}$, $P_{c(\frac{3}{2},\frac{1}{2})}$
and $P_{c(\frac{3}{2},\frac{3}{2})}$ are set. We find the equation
for $|\frac{1}{2},\frac{1}{2}\rangle$ has a solution for $\lambda$
falling into the reasonable range. It means that $P_c(4312)$is
maybe a molecular state of $\Sigma_c \bar D$. The decay width of
$P_{c(\frac{1}{2},\frac{1}{2})}\to J/\psi p$ is calculated within
this framework and we obtain it as about 3.66 MeV.

It is noted, we ignore the couple
channel effects in the Bethe-Salpeter. We also note that if the couple channel
interaction between $\Lambda_c \bar D$ and $\Sigma_c \bar D$ is taken
into account, just as
the authors of Ref.\cite{Shen:2017ayv} did, a bound state of
$\Lambda_c \bar D$ may exist via the coupled channel with
$\Sigma_c \bar D$ (I=$\frac{1}{2}$). In other words, there
is a $\Lambda_c \bar D$ component in the physical state of
$P_{c(\frac{1}{2},\frac{1}{2})}$.

In this work, we
study  $\Lambda_b  B$ and $\Sigma_b  B$ systems and
solve the B-S equations for  $\Sigma_b  B$ and
$\Lambda_b  B$.  Our conclusion is that the bound state
$P_{b(\frac{1}{2},\frac{1}{2})}$ is still a promising pentaquark.
The partial width $P_{b(\frac{1}{2},\frac{1}{2})}\to
\Upsilon p$ is about 1.06 keV at $\Delta E=20$ MeV, which will be
checked by the future experiments.

Within the B-S framework, we systematically investigate the
molecular structure of pentaquarks. We pay a special attention to
$P_c(4312)$ because it is experimentally well measured. From that
study, we have accumulated valuable knowledge on probable
molecular structure of pentaquarks which can be applied to the
future research.  Definitely, the discovery of pentaquarks opens a
window for understanding the quark model established by Gell-Mann
and several other predecessors. Deeper study on their structure
and concerned effective interaction which binds the ingredients to
form a molecule would greatly enrich our theoretical asset. So we
will continue to do research along the line.

\section*{Acknowledgement}
This work is supported by the National Natural Science Foundation
of China (NNSFC) under Contract No. 11375128, , 11675082, 11735010
and 11975165.. We would like to
 thank professors Bing-Song Zou, Xiang Liu and Yu-Ming Wang, as well as Dr. Zhen-Yang Wang for their
 suggestions and useful discussions.

\appendix
\section{The effective interactions}

The effective interactions can be found
in\cite{Ronchen:2012eg,Shen:2016tzq,He:2017aps}
\begin{eqnarray}
&&\mathcal{L}_{\mathcal{P}\mathcal{P}\rho}=ig_{\mathcal{P}\mathcal{P}\rho}\phi_\mathcal{P}{\rho}^\mu\cdot{\tau}
\partial_\mu \phi_\mathcal{P}+c.c.,\\&&
\mathcal{L}_{\mathcal{PPV}}=ig_{\mathcal{PPV}}\phi(x)_\mathcal{P}\partial_\mu\phi(x)_\mathcal{P}\phi(x)_\mathcal{V}^\mu+c.c.,
\\&&\mathcal{L}_{\mathcal{BB}\rho}=-g_{\mathcal{BB}\rho}\bar \psi_\mathcal{B}(\gamma^\mu-\frac{\kappa_{\mathcal{BB}\rho}}{2m_\mathcal{B}}\sigma^{\mu\nu}\partial_\nu){\rho}^\mu\cdot{\tau}
 \psi_\mathcal{B},\\&&\mathcal{L}_{\mathcal{BB}\omega}=-g_{\mathcal{BB}\omega}\bar \psi_\mathcal{B}(\gamma^\mu-\frac{\kappa_{\mathcal{BB}\omega}}{2m_\mathcal{B}}\sigma^{\mu\nu}\partial_\nu)\omega^\mu
 \psi_\mathcal{B},
\\&&
\mathcal{L}_{\mathcal{VVP}}=ig_{\mathcal{VVP}}\varepsilon_{\mu,\nu,\alpha,\beta}\partial^\mu\phi_\mathcal{V}(x)^\nu\partial_\alpha\phi_\mathcal{V}(x)^\beta\phi_\mathcal{P}(x)+c.c.,
\\&&\mathcal{L}_{\mathcal{BBP}}=ig_{\mathcal{BBP}}\bar
\psi_\mathcal{B}\gamma^5\psi_\mathcal{B},
\end{eqnarray}
where $c.c.$ is the complex conjugate term, $\tau$ is pauli matrix for $I=\frac{1}{2}$ and $\tau_x=
\frac{1}{\sqrt{2}}\left(\begin{array}{ccc}
        0 &1 &0 \\
         1 &0&1\\
         0 & 1 & 0
      \end{array}\right)$,
$\tau_y=\frac{1}{\sqrt{2}}\left(\begin{array}{ccc}
        0 &-i &0 \\
         i &0&-i\\
         0 & i & 0
      \end{array}\right)$ and $\tau_z=\frac{1}{\sqrt{2}}\left(\begin{array}{ccc}
        1 &0 &0 \\
         0 &0&0\\
         0 &0 & -1
      \end{array}\right)$ for $I=1$. when
$\phi_\mathcal{P}=\left(\begin{array}{c}
      D^0 \\
        D^+
      \end{array}\right)$ and $\psi_\mathcal{B}=\left(\begin{array}{c}
      \Sigma_{c}^{++} \\
        \Sigma_{c}^+\\
        \Sigma_{c}^0
      \end{array}\right)$ the effective
interactions are consistent with those in Ref\cite{Chen:2019asm}.

\section{The isospin factors in the kernel}

To gain the characteristic hadronic property of the pentaquark, one needs to project the bound states on the vacuum via the field operators
$\mathcal{B}_1$, $\mathcal{B}_2$, $\mathcal{M}_1$ and
$\mathcal{M}_2$ and
\begin{eqnarray} \label{4-dim-BS21}
\langle0|T\mathcal{B}_i(x_1)\mathcal{M}_j(x_2)|P\rangle_{I,I_3}=C^{ij}_{(I,I_3)}\chi^I_P(x_1,x_2)
,\end{eqnarray} where $\chi^I_P(x_1,x_2)$ is the B-S wave function
for the bound state with isospin $I$. The isospin coefficients
$C^{22}_{(\frac{1}{2},\frac{1}{2})}$ for $\Lambda_cD$ bound state
is 1, the isospin coefficients for $\Sigma_cD$ bound states are
\begin{eqnarray} \label{4-dim-BS11}
C^{11}_{(\frac{1}{2},\frac{1}{2})}=\sqrt{\frac{2}{3}},\,\,C^{22}_{(\frac{1}{2},\frac{1}{2})}=-\sqrt{\frac{1}{3}}
,\,\,C^{11}_{(\frac{3}{2},\frac{1}{2})}=\sqrt{\frac{2}{3}},\,\,C^{22}_{(\frac{3}{2},\frac{1}{2})}=\sqrt{\frac{1}{3}}
,\,\,C^{12}_{(\frac{3}{2},\frac{3}{2})}=1.
\end{eqnarray}

Then corresponding B-S equation was deduced in
Ref.\cite{Wang:2019krq} as
\begin{eqnarray} \label{4-dim-BS5}
C^{ij}_{(I,I_3)}\chi^I_{{P}}({ p})
=S_\mathcal{B}(p_1)\int{d^4{q}\over(2\pi)^3}\,K^{ij,lk}(P,p,q)C^{lk}_{(I,I_3)}\chi^I_{_{P}}(q)S_\mathcal{M}(p_2)\,,
\end{eqnarray}
where $K^{ij,lk}(P,p,q)$ is still the kernel and its superscripts
$ij$ and $lk$ denote the initial and final  components.

For $\Lambda_c\bar D$
\begin{eqnarray} \label{4-dim-BSc1}
\chi_{{P}}({ p})
=S_\mathcal{B}(p_1)\int{d^4{q}\over(2\pi)^3}\,K^{22,22}\chi_{_{P}}(q)S_\mathcal{M}(p_2)\,,
\end{eqnarray}
For $I=\frac{1}{2}\,,I_z=\frac{1}{2}$ state $\Sigma_c\bar D$  if
the components are $\Sigma_c^{++}D^-$
\begin{eqnarray} \label{4-dim-BSc2}
\chi_{{P}}({ p})
=S_\mathcal{B}(p_1)\int{d^4{q}\over(2\pi)^3}\,(-K^{11,11}-\frac{1}{\sqrt{2}}K^{11,22})\chi_{_{P}}(q)S_\mathcal{M}(p_2)\,,
\end{eqnarray}
if the components are $\Sigma_c^{+}\bar D^0$
\begin{eqnarray} \label{4-dim-BSc3}
\chi_{{P}}({ p})
=S_\mathcal{B}(p_1)\int{d^4{q}\over(2\pi)^3}\,(K^{22,22}-{\sqrt{2}}K^{22,11})\chi_{_{P}}(q)S_\mathcal{M}(p_2)\,,
\end{eqnarray}
so \begin{eqnarray} \label{4-dim-BSc4} \chi_{{P}}({ p})
=S_\mathcal{B}(p_1)\int{d^4{q}\over(2\pi)^3}\,(-\frac{2}{3}K^{11,11}-\frac{\sqrt{2}}{3}K^{11,22}+
\frac{1}{3}K^{22,22}-\frac{\sqrt{2}}{3}K^{22,11})\chi_{_{P}}(q)S_\mathcal{M}(p_2)\,,
\end{eqnarray}

For $I=\frac{3}{2}\,,I_z=\frac{1}{2}$  $\Sigma_c\bar D$ state if
the components are $\Sigma_c^{++}D^-$
\begin{eqnarray} \label{4-dim-BSc5}
\chi_{{P}}({ p})
=S_\mathcal{B}(p_1)\int{d^4{q}\over(2\pi)^3}\,(-K^{11,11}+{\sqrt{2}}K^{11,22})\chi_{_{P}}(q)S_\mathcal{M}(p_2)\,,
\end{eqnarray}
if the components are $\Sigma_c^{+}\bar D^0$
\begin{eqnarray} \label{4-dim-BSc6}
\chi_{{P}}({ p})
=S_\mathcal{B}(p_1)\int{d^4{q}\over(2\pi)^3}\,(K^{22,22}+\frac{1}{\sqrt{2}}K^{22,11})\chi_{_{P}}(q)S_\mathcal{M}(p_2)\,,
\end{eqnarray}
so \begin{eqnarray} \label{4-dim-BSc7} \chi_{{P}}({ p})
=S_\mathcal{B}(p_1)\int{d^4{q}\over(2\pi)^3}\,(-\frac{1}{3}K^{11,11}+\frac{\sqrt{2}}{3}K^{11,22}+
\frac{2}{3}K^{22,22}+\frac{\sqrt{2}}{3}K^{22,11})\chi_{_{P}}(q)S_\mathcal{M}(p_2)\,.
\end{eqnarray}

The sign ``-" before $K^{11,11}$ in Eq. (\ref{4-dim-BSc2}) and
(\ref{4-dim-BSc5}) comes from the interactions in Appendix A.  For
$\Lambda_c\bar D$ state the two components interact only by
exchanging $\omega$. However $\omega$ and $\rho$ can contribute to
the $\Sigma_c\bar D$ state. One also has
$K^{11,11}(\omega)=K^{22,22}(\omega)$,
$K^{11,22}(\omega)=K^{22,11}(\omega)=0$,
$K^{11,22}(\rho)=K^{22,11}(\rho)=\sqrt{2}K^{11,11}(\rho)$ and
$K^{22,22}(\rho)=0$. In the Eqs. (\ref{4-dim-BSc4}) and
(\ref{4-dim-BSc7}) $K^{11,11}$, $K^{11,22}$ and $K^{22,11}$ can be
changed into $K^{11,11}$ and then the coefficient of $K^{11,11}$
is just the isospin factor $C_{I,I_z}:$
$C_{\frac{1}{2},\frac{1}{2}}=1,-2$ for $\omega$ and $\rho$,
$C_{\frac{3}{2},\frac{1}{2}}=C_{\frac{3}{2},\frac{3}{2}}=1,1$ for
$\omega$ and $\rho$.

\section{The coupled equation of $f_1(|\mathbf{p}_T|)$ and
$f_2(|\mathbf{p}_T|)$ after integrating over $p_l$ and some formulas for azimuthal integration}
\begin{eqnarray} \label{couple equation12}
&&f_1(|\mathbf{p}_T|)=-\int\frac{d^3\mathbf{q}_T}{(2\pi)^3}\frac{C_{I,I_z}g_{\mathcal{MMV}}g_{\mathcal{BBV}}
F^2(k,m_V)}
{2\omega_1(M+\omega_1+\omega_2)(M+\omega_1-\omega_2)}\nonumber\\&&\{\frac{{\mathbf{p}_T\cdot
\mathbf{q}_T} + {\mathbf{p}_T}^2 + 2( {m_1} - {\omega_1} ) (
M+{\omega_1}
 ) +{({\mathbf{p}_T}^2 -{\mathbf{p}_T\cdot \mathbf{q}_T}  )
( {\mathbf{p}_T}^2 -{\mathbf{q}_T}^2) }
 / {{m_V}^2}}{-(\mathbf{p}_T-\mathbf{q}_T)^2-m_V^2}f_1(|\mathbf{q}_T|)+\nonumber\\&&\frac{\frac{( {\mathbf{p}_T}^2 -{\mathbf{q}_T}^2)
   (\mathbf{p}_T\cdot \mathbf{q}_T - {\mathbf{q}_T}^2 )
    ( {m_1} - {\omega_l} ) }{{m_V}^2}
  + (m_1-\omega_1)(
{\mathbf{p}_T\cdot \mathbf{q}_T} + {\mathbf{q}_T}^2 )  +
2(M+\omega_1)\mathbf{p}_T\cdot
\mathbf{q}_T}{-(\mathbf{p}_T-\mathbf{q}_T)^2-m_V^2}f_2(|\mathbf{q}_T|)\nonumber\\&&-\frac{\kappa}{4m_B}\frac{-4[
{{\mathbf{p}_T\cdot \mathbf{q}_T}}^2 -
{\mathbf{p}_T}^2{\mathbf{q}_T}^2 +
(m_1-\omega_1)({\mathbf{p}_T\cdot \mathbf{q}_T}- {\mathbf{q}_T}^2
)(\omega_1+M)
    ]}{-(\mathbf{p}_T-\mathbf{q}_T)^2-m_V^2}f_2(|\mathbf{q}_T|)\nonumber\\&&-\frac{\kappa}{4m_B}\frac{4(
{  {\mathbf{p}_T}^2 -\mathbf{p}_T\cdot \mathbf{q}_T}) ( M+\omega_1
)}{-(\mathbf{p}_T-\mathbf{q}_T)^2-m_V^2}f_1(|\mathbf{q}_T|)\}+\nonumber\\&&
\int\frac{d^3\mathbf{q}_T}{(2\pi)^3}\frac{C_{I,I_z}g_{\mathcal{MMV}}g_{\mathcal{BBV}}
F^2(k,m_V)}
{2\omega_2(M+\omega_1-\omega_2)(M-\omega_1-\omega_2)}\nonumber\\&&\{\frac{{\mathbf{p}_T\cdot
\mathbf{q}_T} + {\mathbf{p}_T}^2 + 2( {m_1}- {\omega_2} +
M)\omega_2 +{({\mathbf{p}_T}^2 -{\mathbf{p}_T\cdot \mathbf{q}_T} )
( {\mathbf{p}_T}^2 -{\mathbf{q}_T}^2) }
 / {{m_V}^2}}{-(\mathbf{p}_T-\mathbf{q}_T)^2-m_V^2}f_1(|\mathbf{q}_T|)+\nonumber\\&&\frac{\frac{( {\mathbf{p}_T}^2 -{\mathbf{q}_T}^2)
   (\mathbf{p}_T\cdot \mathbf{q}_T - {\mathbf{q}_T}^2 )
    ( {m_1} - {\omega_2} + M ) }{{m_V}^2}
  + (m_1+M-\omega_2)(
{\mathbf{p}_T\cdot \mathbf{q}_T} + {\mathbf{q}_T}^2 )  +
2\omega_2\mathbf{p}_T\cdot
\mathbf{q}_T}{-(\mathbf{p}_T-\mathbf{q}_T)^2-m_V^2}f_2(|\mathbf{q}_T|)\nonumber\\&&-\frac{\kappa}{4m_B}\frac{-4[
{{\mathbf{p}_T\cdot \mathbf{q}_T}}^2 -
{\mathbf{p}_T}^2{\mathbf{q}_T}^2 + (m_1-\omega_2+M)(
{\mathbf{q}_T}^2 -{\mathbf{p}_T\cdot \mathbf{q}_T} )(-\omega_2)
    ]}{-(\mathbf{p}_T-\mathbf{q}_T)^2-m_V^2}f_2(|\mathbf{q}_T|)\nonumber\\&&-\frac{\kappa}{4m_B}\frac{4(
{\mathbf{p}_T\cdot \mathbf{q}_T} - {\mathbf{p}_T}^2 )
(-\omega_2)}{-(\mathbf{p}_T-\mathbf{q}_T)^2-m_V^2}f_1(|\mathbf{q}_T|)\}.
\end{eqnarray}

\begin{eqnarray} \label{couple equation22}
&&f_2(|\mathbf{p}_T|){\mathbf{p}_T}^2=-\int\frac{d^3\mathbf{q}_T}{(2\pi)^3}\frac{-C_{I,I_z}g_{\mathcal{MMV}}g_{\mathcal{BBV}}
F^2(k,m_V)}
{2\omega_1(M+\omega_1+\omega_2)(M+\omega_1-\omega_2)}\nonumber\\&&\{\frac{-{\mathbf{p}_T}^2(\mathbf{p}_T\cdot
\mathbf{q}_T + {\mathbf{q}_T}^2) + 2{\mathbf{p}_T\cdot
\mathbf{q}_T}( {m_1}+{\omega_1}) ( M+{\omega_1}
 ) +{\mathbf{p}_T}^2\frac{({\mathbf{q}_T}^2 -{\mathbf{p}_T\cdot \mathbf{q}_T}  )
( {\mathbf{p}_T}^2 -{\mathbf{q}_T}^2) }
  {{{m_V}}^2}}{-(\mathbf{p}_T-\mathbf{q}_T)^2-m_V^2}f_2(|\mathbf{q}_T|)+\nonumber\\&&\frac{\frac{(
   {\mathbf{p}_T}^2 -{\mathbf{q}_T}^2) (\mathbf{p}_T\cdot \mathbf{q}_T - {\mathbf{p}_T}^2 )
    ( -{m_1}- {\omega_1}  ) }{{{m_V}}^2}
  + (m_1-\omega_1)(
{\mathbf{p}_T\cdot \mathbf{q}_T} + {\mathbf{p}_T}^2 )  -
2M{\mathbf{p}_T}^2+2\omega_1\mathbf{p}_T\cdot
\mathbf{q}_T}{-(\mathbf{p}_T-\mathbf{q}_T)^2-m_V^2}f_1(|\mathbf{q}_T|)\nonumber\\&&-\frac{\kappa}{4m_B}\frac{4[(
{M} +{\omega_1} ) {\mathbf{p}_T\cdot \mathbf{q}_T}{\mathbf{p}_T}^2
-
 {\mathbf{p}_T\cdot \mathbf{q}_T}^2( {m_1} + {\omega_1} )  +
  {\mathbf{p}_T}^2{\mathbf{q}_T}^2( {m_1} - M
     )]}{-(\mathbf{p}_T-\mathbf{q}_T)^2-m_V^2}f_2(|\mathbf{q}_T|)\nonumber\\&&-\frac{\kappa}{4m_B}\frac{4(
{\mathbf{p}_T\cdot \mathbf{q}_T} - {\mathbf{p}_T}^2 ) (-\omega_1 -
M
)(m_1+\omega_1)}{-(\mathbf{p}_T-\mathbf{q}_T)^2-m_V^2}f_1(|\mathbf{q}_T|)\}\nonumber\\&&+\int\frac{d^3\mathbf{q}_T}{(2\pi)^3}\frac{-C_{I,I_z}g_{\mathcal{MMV}}g_{\mathcal{BBV}}
F^2(k,m_V)}
{2\omega_2(M+\omega_1-\omega_2)(M-\omega_1-\omega_2)}\nonumber\\&&\{\frac{-{\mathbf{p}_T}^2(\mathbf{p}_T\cdot
\mathbf{q}_T + {\mathbf{q}_T}^2) + 2{\mathbf{p}_T\cdot
\mathbf{q}_T}( {m_1}+{\omega_2} - M)  \omega_2
  +{\mathbf{p}_T}^2\frac{({\mathbf{q}_T}^2 -{\mathbf{p}_T\cdot \mathbf{q}_T}  )
( {\mathbf{p}_T}^2 -{\mathbf{q}_T}^2) }
  {{{m_V}}^2}}{-(\mathbf{p}_T-\mathbf{q}_T)^2-m_V^2}f_2(|\mathbf{q}_T|)+\nonumber\\&&\frac{\frac{( {\mathbf{p}_T}^2 -{\mathbf{q}_T}^2) (\mathbf{p}_T\cdot \mathbf{q}_T - {\mathbf{p}_T}^2 )
    (  M-{m_1} - {\omega_2}  ) }{{{m_V}}^2}
  + (m_1-M-\omega_2)(
{\mathbf{p}_T\cdot \mathbf{q}_T} + {\mathbf{p}_T}^2 )
+2\omega_2\mathbf{p}_T\cdot
\mathbf{q}_T}{-(\mathbf{p}_T-\mathbf{q}_T)^2-m_V^2}f_1(|\mathbf{q}_T|)\nonumber\\&&-\frac{\kappa}{4m_B}\frac{4[
\omega_2 {\mathbf{p}_T\cdot \mathbf{q}_T}{\mathbf{p}_T}^2 -
 {\mathbf{p}_T\cdot \mathbf{q}_T}^2( {m_1} + {\omega_2}- M )  +
  {\mathbf{p}_T}^2{\mathbf{q}_T}^2( {m_1} - M
     )]}{-(\mathbf{p}_T-\mathbf{q}_T)^2-m_V^2}f_2(|\mathbf{q}_T|)\nonumber\\&&-\frac{\kappa}{4m_B}\frac{4(
{\mathbf{p}_T\cdot \mathbf{q}_T} - {\mathbf{p}_T}^2 ) \omega_2
(M-m_1-\omega_2)}{-(\mathbf{p}_T-\mathbf{q}_T)^2-m_V^2}f_1(|\mathbf{q}_T|)\}.
\end{eqnarray}

Since $d^3\mathbf{q}_T=\mathbf{q}_T^2{\rm
sin}(\theta)d|\mathbf{q}_T|d\theta d\phi$ and $\mathbf{p}_T\cdot
\mathbf{q}_T=|\mathbf{p}_T||\mathbf{q}_T|{\rm cos}(\theta)$ one
can carry out the azimuthal integration for Eqs. (\ref{couple
equation12}) and (\ref{couple equation22}) analytically. Some
useful integrations are defined as follow

\begin{eqnarray} \label{azimuthal1}
&&J_0\equiv\int_0^\pi{\rm sin}(\theta)d\theta
\frac{1}{-(\mathbf{p}_T-\mathbf{q}_T)^2-m_V^2}[\frac{\Lambda^2-m^2_V}{\Lambda^2-{(\mathbf{p}_T-\mathbf{q}_T)^2}}]^2
\nonumber\\&&=\int_0^\pi \frac{{\rm
sin}(\theta)d\theta}{-[\mathbf{p}_T^2+\mathbf{q}_T^2-2|\mathbf{p}_T||\mathbf{q}_T|{\rm
cos}(\theta)]-m_V^2}\{\frac{\Lambda^2-m^2_V}{\Lambda^2-{[\mathbf{p}_T^2+\mathbf{q}_T^2-2|\mathbf{p}_T||\mathbf{q}_T|{\rm
cos}(\theta)]}}\}^2\nonumber\\&&=-\frac{2(m_V^2-\Lambda^2)}{[(|\mathbf{p}_T|-|\mathbf{q}_T|)^2+\Lambda^2][(|\mathbf{p}_T|+|\mathbf{q}_T|)^2
+\Lambda^2]}\nonumber\\&&+\frac{1}{2|\mathbf{p}_T||\mathbf{q}_T|}\{{\rm
Ln}[\frac{(|\mathbf{p}_T|+|\mathbf{q}_T|)^2+\Lambda^2}{(|\mathbf{p}_T|-|\mathbf{q}_T|)^2+\Lambda^2}]-{\rm
Ln}[\frac{(|\mathbf{p}_T|+
|\mathbf{q}_T|)^2+m_V^2}{(|\mathbf{p}_T|-|\mathbf{q}_T|)^2+m_V^2}]\},
\end{eqnarray}

\begin{eqnarray} \label{azimuthal2}
&&J_1\equiv\int_0^\pi {\rm
sin}(\theta)d\theta\frac{\mathbf{p}_T\cdot
\mathbf{q}_T}{-(\mathbf{p}_T-\mathbf{q}_T)^2-m_V^2}[\frac{\Lambda^2-m^2_V}{\Lambda^2-{(\mathbf{p}_T-\mathbf{q}_T)^2}}]^2
\nonumber\\&&=\int_0^\pi \frac{|\mathbf{p}_T||\mathbf{q}_T|{\rm
cos}(\theta){\rm
sin}(\theta)d\theta}{-[\mathbf{p}_T^2+\mathbf{q}_T^2-2|\mathbf{p}_T||\mathbf{q}_T|{\rm
cos}(\theta)]-m_V^2}\{\frac{\Lambda^2-m^2_V}{\Lambda^2-{[\mathbf{p}_T^2+\mathbf{q}_T^2-2|\mathbf{p}_T||\mathbf{q}_T|{\rm
cos}(\theta)]}}\}^2\nonumber\\&&=-\frac{(m_V^2-\Lambda^2)(|\mathbf{p}_T|^2+|\mathbf{q}_T|^2
+\Lambda^2)}{[(|\mathbf{p}_T|-|\mathbf{q}_T|)^2+\Lambda^2][(|\mathbf{p}_T|+|\mathbf{q}_T|)^2
+\Lambda^2]}\nonumber\\&&+\frac{(|\mathbf{p}_T|^2+|\mathbf{q}_T|^2
+m_V^2)}{4|\mathbf{p}_T||\mathbf{q}_T|}\{Ln[\frac{(|\mathbf{p}_T|+|\mathbf{q}_T|)^2+\Lambda^2}{(|\mathbf{p}_T|-|\mathbf{q}_T|)^2+\Lambda^2}]-Ln[\frac{(|\mathbf{p}_T|+
|\mathbf{q}_T|)^2+m_V^2}{(|\mathbf{p}_T|-|\mathbf{q}_T|)^2+m_V^2}]\},
\end{eqnarray}

\begin{eqnarray} \label{azimuthal3}
&&J_2\equiv\int_0^\pi {\rm
sin}(\theta)d\theta\frac{(\mathbf{p}_T\cdot
\mathbf{q}_T)^2}{-(\mathbf{p}_T-\mathbf{q}_T)^2-m_V^2}[\frac{\Lambda^2-m^2_V}{\Lambda^2-{(\mathbf{p}_T-\mathbf{q}_T)^2}}]^2
\nonumber\\&&=\int_0^\pi
\frac{|\mathbf{p}_T|^2|\mathbf{q}_T|^2{\rm cos}^2(\theta){\rm
sin}(\theta)d\theta}{-[\mathbf{p}_T^2+\mathbf{q}_T^2-2|\mathbf{p}_T||\mathbf{q}_T|{\rm
cos}(\theta)]-m_V^2}\{\frac{\Lambda^2-m^2_V}{\Lambda^2-{[\mathbf{p}_T^2+\mathbf{q}_T^2-2|\mathbf{p}_T||\mathbf{q}_T|{\rm
cos}(\theta)]}}\}^2\nonumber\\&&=-\frac{(m_V^2-\Lambda^2)(|\mathbf{p}_T|^2+|\mathbf{q}_T|^2
+\Lambda^2)^2}{2[(|\mathbf{p}_T|-|\mathbf{q}_T|)^2+\Lambda^2][(|\mathbf{p}_T|+|\mathbf{q}_T|)^2
+\Lambda^2]}\nonumber\\&&+\frac{1}{8|\mathbf{p}_T||\mathbf{q}_T|}\{(|\mathbf{p}_T|^2+|\mathbf{q}_T|^2
+2m_V^2-\Lambda^2)(|\mathbf{p}_T|^2+|\mathbf{q}_T|^2
+\Lambda^2){\rm
Ln}[\frac{(|\mathbf{p}_T|+|\mathbf{q}_T|)^2+\Lambda^2}{(|\mathbf{p}_T|-|\mathbf{q}_T|)^2+\Lambda^2}]\nonumber\\&&-(|\mathbf{p}_T|^2+|\mathbf{q}_T|^2
+m_V^2)^2{\rm Ln}[\frac{(|\mathbf{p}_T|+
|\mathbf{q}_T|)^2+m_V^2}{(|\mathbf{p}_T|-|\mathbf{q}_T|)^2+m_V^2}]\}.
\end{eqnarray}

\begin{eqnarray} \label{couple equation13}
&&A_{11}(\mathbf{p_T},\mathbf{q_T})=\frac{-\mathbf{q}_T^2}{(2\pi)^2}\frac{C_{I,I_z}g_{\mathcal{MMV}}g_{\mathcal{BBV}}}
{2\omega_1(M+\omega_1+\omega_2)(M+\omega_1-\omega_2)}\nonumber\\&&\{{[J_1
+ {\mathbf{p}_T}^2J_0 + 2( {m_1} - {\omega_1} ) ( M+{\omega_1}
 )J_0 +{({\mathbf{p}_T}^2J_0 -J_1  )
( {\mathbf{p}_T}^2 -{\mathbf{q}_T}^2) }
 / {{m_V}^2}]}f_1(|\mathbf{q}_T|)\nonumber\\&&-\frac{\kappa}{4m_B}{[4(
{  {\mathbf{p}_T}^2J_0 -J_1}) ( M{\eta_2}+\omega_1
)]}f_1(|\mathbf{q}_T|)\}+\nonumber\\&&
\frac{\mathbf{q}_T^2}{(2\pi)^2}\frac{C_{I,I_z}g_{\mathcal{MMV}}g_{\mathcal{BBV}}
}
{2\omega_2(M+\omega_1-\omega_2)(M-\omega_1-\omega_2)}\nonumber\\&&\{{[J_1
+ {\mathbf{p}_T}^2J_0 + 2( {m_1}- {\omega_2} + M)\omega_2J_0
+{({\mathbf{p}_T}^2J_0 -J_1 ) ( {\mathbf{p}_T}^2
-{\mathbf{q}_T}^2) }
 / {{m_V}^2}]}f_1(|\mathbf{q}_T|)\nonumber\\&&-\frac{\kappa}{4m_B}{[4(
J_1 - {\mathbf{p}_T}^2J_0 ) (-\omega_2)]}f_1(|\mathbf{q}_T|)\}.
\end{eqnarray}

\begin{eqnarray} \label{couple equation14}
&&A_{12}(\mathbf{p_T},\mathbf{q_T})=\frac{-\mathbf{q}_T^2}{(2\pi)^2}\frac{C_{I,I_z}g_{\mathcal{MMV}}g_{\mathcal{BBV}}}
{2\omega_1(M+\omega_1+\omega_2)(M+\omega_1-\omega_2)}\nonumber\\&&\{{[\frac{(
{\mathbf{p}_T}^2 -{\mathbf{q}_T}^2)
   (J_1 - {\mathbf{q}_T}^2J_0 )
    ( {m_1} - {\omega_l} ) }{{m_V}^2}
  + (m_1-\omega_1)(
J_1 + {\mathbf{q}_T}^2J_0 )  +
2(M+\omega_1)J_1]}f_2(|\mathbf{q}_T|)\nonumber\\&&+\frac{\kappa}{m_B}{[
J_2 - {\mathbf{p}_T}^2{\mathbf{q}_T}^2J_0 + (m_1-\omega_1)(J_1-
{\mathbf{q}_T}^2J_0 )(\omega_1+M)
    ]}f_2(|\mathbf{q}_T|)\}+\nonumber\\&&
\frac{\mathbf{q}_T^2}{(2\pi)^2}\frac{C_{I,I_z}g_{\mathcal{MMV}}g_{\mathcal{BBV}}
}
{2\omega_2(M+\omega_1-\omega_2)(M-\omega_1-\omega_2)}\nonumber\\&&\{{[\frac{(
{\mathbf{p}_T}^2 -{\mathbf{q}_T}^2)
   (J_1 - {\mathbf{q}_T}^2J_0 )
    ( {m_1} - {\omega_2} + M ) }{{m_V}^2}
  + (m_1+M-\omega_2)(
J_1 + {\mathbf{q}_T}^2J_0 )  +
2\omega_2J_1]}f_2(|\mathbf{q}_T|)\nonumber\\&&+\frac{\kappa}{m_B}{[
J_2 - {\mathbf{p}_T}^2{\mathbf{q}_T}^2J_0 + (m_1-\omega_2+M)(
{\mathbf{q}_T}^2J_0 -J_1 )(-\omega_2)
    ]}f_2(|\mathbf{q}_T|)\}.
\end{eqnarray}

\begin{eqnarray} \label{couple equation23}
&&A_{21}(\mathbf{p_T},\mathbf{q_T}){\mathbf{p}_T}^2=\frac{\mathbf{q}_T^2}{(2\pi)^2}\frac{C_{I,I_z}g_{\mathcal{MMV}}g_{\mathcal{BBV}}
}
{2\omega_1(M+\omega_1+\omega_2)(M+\omega_1-\omega_2)}\nonumber\\&&\{[{\frac{(
   {\mathbf{p}_T}^2 -{\mathbf{q}_T}^2) (J_1 - {\mathbf{p}_T}^2J_0 )
    ( -{m_1}- {\omega_1}  ) }{{{m_V}}^2}
  + (m_1-\omega_1)(
J_1 + {\mathbf{p}_T}^2J_0 )  -
2M{\mathbf{p}_T}^2J_0+2\omega_1J_1}]f_1(|\mathbf{q}_T|)\nonumber\\&&-\frac{\kappa}{4m_B}[\frac{4(
J_1 - {\mathbf{p}_T}^2J_0 ) (-\omega_1 - M
)(m_1+\omega_1)}{-(\mathbf{p}_T-\mathbf{q}_T)^2-m_V^2}]f_1(|\mathbf{q}_T|)\}\nonumber\\&&+\frac{\mathbf{q}_T^2}{(2\pi)^2}\frac{C_{I,I_z}g_{\mathcal{MMV}}g_{\mathcal{BBV}}}
{2\omega_2(M+\omega_1-\omega_2)(M-\omega_1-\omega_2)}\nonumber\\&&\{[{\frac{(
{\mathbf{p}_T}^2 -{\mathbf{q}_T}^2) (J_1 - {\mathbf{p}_T}^2J_0 )
    (  M-{m_1} - {\omega_2}  ) }{{{m_V}}^2}
  + (m_1-M-\omega_2)(
J_1 + {\mathbf{p}_T}^2J_0 )
+2\omega_2J_1}]f_1(|\mathbf{q}_T|)\nonumber\\&&-\frac{\kappa}{4m_B}{[4(
J_1 - {\mathbf{p}_T}^2J_0 ) \omega_2
(M-m_1-\omega_2)]}f_1(|\mathbf{q}_T|)\}.
\end{eqnarray}

\begin{eqnarray} \label{couple equation24}
&&A_{22}(\mathbf{p_T},\mathbf{q_T}){\mathbf{p}_T}^2=\frac{\mathbf{q}_T^2}{(2\pi)^2}\frac{C_{I,I_z}g_{\mathcal{MMV}}g_{\mathcal{BBV}}}
{2\omega_1(M+\omega_1+\omega_2)(M+\omega_1-\omega_2)}\nonumber\\&&\{[{-{\mathbf{p}_T}^2(J_1
+ {\mathbf{q}_T}^2J_0) + 2J_1( {m_1}+{\omega_1}) ( M+{\omega_1}
 ) +{\mathbf{p}_T}^2\frac{({\mathbf{q}_T}^2J_0 -J_1  )
( {\mathbf{p}_T}^2 -{\mathbf{q}_T}^2) }
  {{{m_V}}^2}}]f_2(|\mathbf{q}_T|)\nonumber\\&&-\frac{\kappa}{4m_B}{4[(
{M} +{\omega_1} ){\mathbf{p}_T}^2 J_1 - ( {m_1} + {\omega_1} )J_2
+
  {\mathbf{p}_T}^2{\mathbf{q}_T}^2( {m_1} - M
     )J_0]}f_2(|\mathbf{q}_T|)\}\nonumber\\&&+\frac{\mathbf{q}_T^2}{(2\pi)^2}\frac{C_{I,I_z}g_{\mathcal{MMV}}g_{\mathcal{BBV}}}
{2\omega_2(M+\omega_1-\omega_2)(M-\omega_1-\omega_2)}\nonumber\\&&\{[{-{\mathbf{p}_T}^2(J_1
+ {\mathbf{q}_T}^2J_0) + 2( {m_1}+{\omega_2} - M)  \omega_2J_1
  +{\mathbf{p}_T}^2\frac{({\mathbf{q}_T}^2J_0 -J_1  )
( {\mathbf{p}_T}^2 -{\mathbf{q}_T}^2) }
  {{{m_V}}^2}}]f_2(|\mathbf{q}_T|)\nonumber\\&&-\frac{\kappa}{4m_B}{4[
\omega_2 {\mathbf{p}_T}^2J_1 - ( {m_1} + {\omega_2}- M )J_2  +
  {\mathbf{p}_T}^2{\mathbf{q}_T}^2( {m_1} - M
     )J_0]}f_2(|\mathbf{q}_T|)\}.
\end{eqnarray}

\end{document}